\documentclass[manuscript]{acmart}

\usepackage[utf8]{inputenc}
\usepackage{subcaption}
\usepackage{mathtools} 
\usepackage{fancyvrb}
\usepackage{listings}
\usepackage{enumitem}
\usepackage{booktabs}
\usepackage{threeparttable}
\usepackage{textcomp}
\usepackage{multirow}
\usepackage{booktabs}
\usepackage{placeins} 
\usepackage{cuted}    
\usepackage{capt-of}  

\usepackage[toc,page]{appendix}

\AtBeginDocument{%
  }

\begin{document}

\title{Trinketry: Materializing Elements to Anchor Traceable Design Histories}


\author{Wen-Fan Wang}
\authornote{Both authors contributed equally.}
\email{vann@cmlab.csie.ntu.edu.tw}
\orcid{0009-0001-1050-1170}
\affiliation{%
  \institution{National Taiwan University}
  \city{Taipei}
  \country{Taiwan}
}

\author{Yi-Ting Chiu}
\authornotemark[1]
\email{R13922018@csie.ntu.edu.tw}
\orcid{0009-0004-0937-2705}
\affiliation{%
  \institution{National Taiwan University}
  \city{Taipei}
  \country{Taiwan}
}

\author{You-Yi Hsieh}
\authornotemark
\email{b11705024@ntu.edu.tw}
\orcid{0009-0001-3046-6173}
\affiliation{%
  \institution{National Taiwan University}
  \city{Taipei}
  \country{Taiwan}
}

\author{Bing-Yu Chen}
\email{robin@ntu.edu.tw}
\orcid{0000-0003-0169-7682}
\affiliation{%
  \institution{National Taiwan University}
  \city{Taipei}
  \country{Taiwan}
  }

\author{Max Kreminski}
\email{maxkreminski@gmail.com}
\orcid{0009-0002-6268-4033}
\affiliation{%
  \institution{Cornell Tech}
  \city{New York, New York}
  \country{USA}
  }

\renewcommand{\shortauthors}{Wang Chiu Hsieh Chen Kreminski}


\begin{abstract}

Generative tools enable visual designers to rapidly explore diverse design directions---but as flawed generated outputs proliferate, designers may lose track of promising ideas and the context of their development. We introduce \emph{trinkets}: user-preserved design elements that function as reusable materials, retrieval queries, and navigational anchors for tracing how ideas emerge, transform, and recur throughout design history. We implement this concept in Trinketry, a visual exploration canvas. In a within-subjects study with 12 concept designers, Trinketry supported focused element-level exploration and greater perceived control than a simple branching image generation history baseline. A follow-up study with six returning designers examined delayed design history sensemaking, continuation of work, and reconstruction of others’ processes. Compared with whole-image histories, Trinketry supported more precise reconstruction of process structure, turning points, and goals, while helping designers recover overlooked materials and continue exploration. However, opportunities remain to augment trace representations with explicit statements of design rationale.

\end{abstract}



\begin{teaserfigure}
  \centering
  \includegraphics[width=1\textwidth]{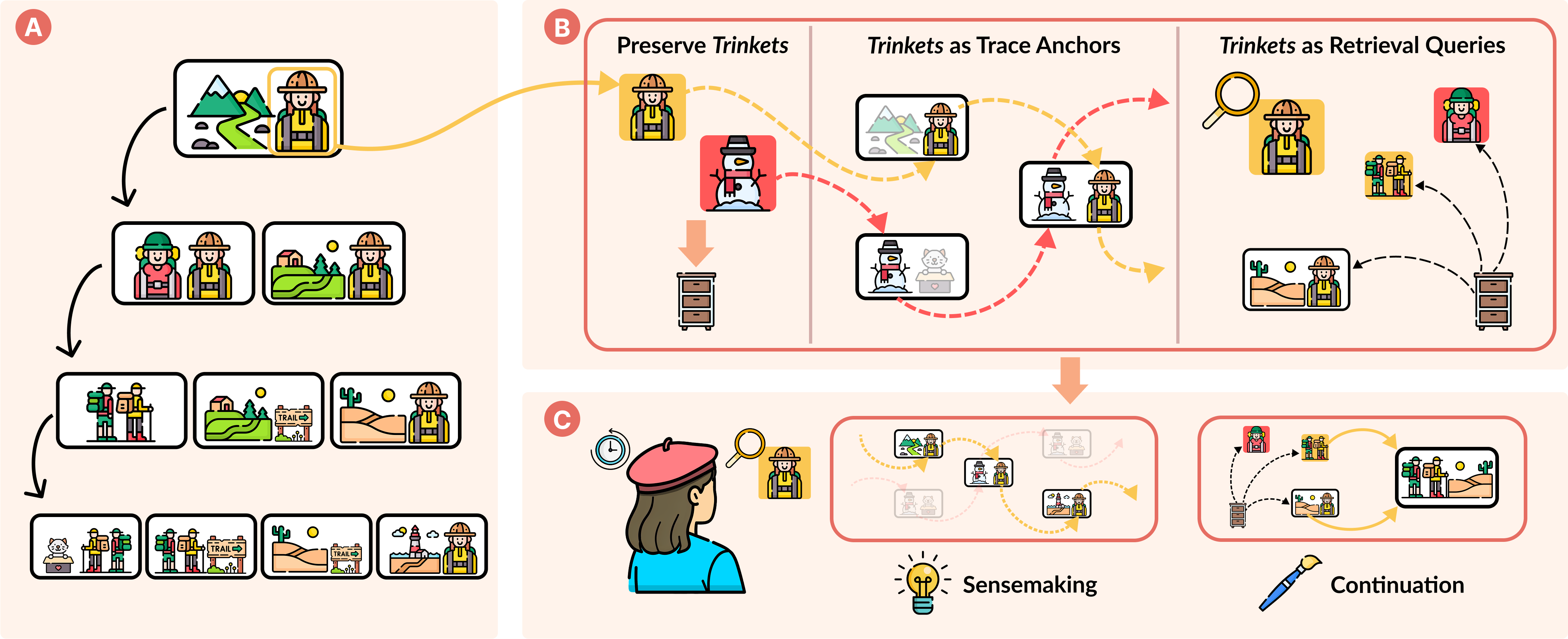}
  \caption{(A) As designers generate and refine images during AI-assisted visual exploration, promising elements can become scattered across a growing collection of intermediate outputs. (B) Trinketry enables designers to extract and preserve these elements as \emph{trinkets}, which can serve as reusable materials for recombination, trace anchors for navigating element provenance, and retrieval queries for surfacing related past materials. (C) When designers revisit their exploration histories after a delay, preserved trinkets provide entry points for making sense of past actions and continuing exploration.}
  \Description[Three-panel overview of scattered design elements, trinket-centered access, and delayed reuse.]{Three-panel conceptual overview showing how Trinketry turns promising elements in generative exploration into persistent trinkets for later access and reuse. Panel A shows a growing sequence of generated image groups connected by arrows. A recurring hiker character appears in several otherwise different images and branches, illustrating how a promising element can persist across outputs while becoming distributed through the exploration history. Panel B shows three roles for preserved trinkets. In “Preserve Trinkets,” individual elements such as the hiker and a snowman are separated from generated images and saved to a drawer. In “Trinkets as Trace Anchors,” preserved elements connect multiple image states through dashed provenance paths, exposing where an element came from and where it later appeared. In “Trinkets as Retrieval Queries,” a preserved hiker trinket is used with a search icon to surface related prior images and trinkets beyond its immediate provenance path. Panel C depicts a designer returning after a delay, indicated by a clock, with a previously preserved hiker trinket available as an entry point. One example shows the designer using connected provenance relations to make sense of the earlier process; another shows preserved and retrieved materials feeding into a new combined image, representing continuation of the prior exploration.}
  \label{fig:teaser}
\end{teaserfigure}

\maketitle

\section{INTRODUCTION}
\label{sec:introduction}

Concept designers shape the visual and spatial worlds of games, animation, film, and television~\cite{bigbadWorld2015, randomguidebook2023, artofgame2008}. Increasingly, they use AI-assisted visual exploration tools to quickly investigate numerous alternative design directions~\cite{filmhandbook2024, longo2024elaborating, long2024sketchar, AIdeation}. While such workflows enable rapid generation of images, the most useful outcome is often not a complete generated image, but a fragment within it: a visual motif, lighting, or abstract element (such as atmosphere or composition) ~\cite{AIdeation, GenTune, zhou2026collaposer}. Designers carry such visual and textual elements across iterations, revisit earlier fragments, and combine materials from distinct explorations into new directions~\cite{choi2024creativeconnect, shi2025brickify, HistoryPalette}. Prior work has begun to support such element-level practices by enabling designers to collect and recombine text labels~\cite{choi2024creativeconnect}, partial visual elements~\cite{HistoryPalette}, intentions~\cite{choi2026ideablocks}, and other multimodal inputs~\cite{DesignPrompt2024}.
However, repeated generation, refinement, branching, and local editing can quickly produce large numbers of intermediate artifacts~\cite{shen2025ideationweb, GenTune, Expandora}, scattering valuable fragments across otherwise disposable outputs and making them difficult to revisit and recover.

Existing tools help manage this growing history through branching interfaces that expose temporal lineage or provenance relationships between images~\cite{DesignTrace, Expandora}, or through spatial grouping and arrangement on a canvas~\cite{ImaginationVellum}. These representations make the evolution of generated images and exploration branches visible, but most still treat images, branches, or canvas states as the primary units of organization and provenance. A design idea may originate in one image, transform across several branches, disappear, and later resurface elsewhere, making its trajectory difficult to capture through image-level provenance alone. As exploration accumulates, designers may therefore be able to see which images were produced and how they are connected, while finding it harder to reconstruct where a particular idea originated, how it changed, where it was reused or recombined, and when it disappeared or resurfaced. In this work, we investigate how user-preserved design elements can become persistent trace anchors for both continuing exploration and reconstructing how ideas evolve over time.

We conducted a formative study with 4 professional concept designers and found that designers’ existing workflows provided limited support for following promising elements once they became distributed across generations. Designers described losing continuity when generated results drifted from desired elements or compositions, and difficulty reconstructing how earlier interactions led to later outcomes. They wanted to revisit meaningful prior states, compare related alternatives, and recover elements that had been lost or transformed. These findings motivated us to treat designer-selected elements as persistent objects that structure exploration history, leading to three design goals: (DG1) support preserving and recombining promising elements across iterations; (DG2) support tracing how promising elements evolve across exploration; and (DG3) recovering relevant prior material for comparison and continued exploration.

Guided by these findings, we introduce \emph{trinkets}: user-preserved design elements that materialize promising ideas discovered during generative exploration. A trinket represents something the designer has deliberately chosen to carry forward and may take the form of an image, selected image region, text concept, or image–text composite. We instantiate this concept in Trinketry, a visual exploration canvas in which trinkets serve three interconnected roles: (1) reusable materials that can be recombined in subsequent generations; (2) trace anchors for navigating where an element originated and how it was reused or transformed; and (3) retrieval queries for surfacing semantically related materials elsewhere in the exploration history. 
By structuring access to prior work around these designer-selected elements, Trinketry turns acts of preservation during exploration into anchors for later reflection, reconstruction, and continuation of the design process.


We evaluated Trinketry through two complementary studies with professional concept designers. In a within-subjects study with 12 designers, we compared Trinketry with a branching image generation history baseline during visual exploration and immediate retrospective tracing. Trinketry supported more focused element-level exploration and greater perceived control, with participants reporting easier reuse, tracing, and retrieval of prior material. After a four-month delay, we conducted an in-depth follow-up with six returning designers to examine the longer-term utility of trace representations for delayed sensemaking, continuation of prior work, and interpretation of other designers' work. Preserved trinkets served as cues for reconstructing process stages, turning points, and material histories, while simple branching image histories provided a smoother global chronological overview. Participants also reused preserved materials to resume earlier explorations. External reviewers could reconstruct visible stages and provenance relations from Trinketry traces, but design rationales remained less recoverable than observable actions and material histories.





In summary, this work makes three contributions:
\begin{itemize}
    \item We introduce \emph{trinkets} and Trinketry, an element-centered representation and system in which user-preserved design elements serve simultaneously as reusable materials, trace anchors, and retrieval queries across generative exploration.

    \item Through a within-subjects study with professional concept designers, we show how element-centered interaction reshapes active exploration by supporting more focused reuse, greater perceived control, and targeted access to prior design material.

    \item Through a four-month delayed follow-up study, we show how preserved elements support delayed process sensemaking and continuation of prior work. We also find that purely action-centered creative activity traces leave design rationales implicit and difficult to reconstruct, suggesting a complementarity between affordances for rationale capture and tracing of material history.
\end{itemize}

\section{RELATED WORK}
\label{sec:relatedwork}

\subsection{Design Elements as Reusable Material}

Visual exploration is central to visual design workflows~\cite{Sourcesofinspiration, randomguidebook2023, wang2026style}, as designers investigate possibilities~\cite{herring2009getting} and gather and organize references~\cite{visualreferences,keller2009collecting,lee2010designing}.
Generative AI has motivated growing research on supporting such exploration. Some systems improve text-to-image generation through interactive prompt refinement~\cite{brade2023promptify, PromptCharm2024, feng2023promptmagician}, while others expand the design space through LLM-assisted ideation~\cite{cai2023designaid, Dreamsheets, poet, wang2023popblends, AIdeation}. GenTune~\cite{GenTune} further connects image elements to prompt structure for traceable element-level editing. Beyond prompting, Inkspire~\cite{Inkspire} supports sketching with real-time analogical inspirations, while FusAIn~\cite{peng2025fusain} and SketchFlex~\cite{lin2025sketchflex} combine sketches with generative outputs on canvas. Across this work, however, element-level interactions primarily support generation and editing, while exploration histories remain largely organized around images, branches, or canvas states.


More recent work treats individual design elements as manipulable and reusable material. GanCollage~\cite{wan2023gancollage}, DesignWeaver~\cite{designweaver}, and DesignPrompt~\cite{DesignPrompt2024} support composing or prompting with reusable visual and multimodal elements, while CreativeConnect~\cite{choi2024creativeconnect} and Creative Blends~\cite{creativeblends} extract semantic concepts from references for blended generation. Brickify~\cite{shi2025brickify} further reifies design elements as reusable tokens for manipulation and recombination, and IdeaBlocks~\cite{choi2026ideablocks} materializes design intents as reusable blocks across explorations. These systems establish the value of element-level interaction, but primarily use elements to construct subsequent generations. Our work instead focuses on how designers deliberately preserve promising elements across iterations. 


\subsection{Tracing Creative Activity}
Interest in the capture, representation, and analysis of \emph{creative activity traces}---temporal data streams reflecting the creative process---is rapidly growing among HCI researchers. Recently, researchers have surveyed the design space of ``trace-aware'' creativity support tools~\cite{TracingCreativity}; convened a CHI workshop on creative activity trace analysis~\cite{HerdingCATs}; and used traces to make sense of human-AI influence dynamics in creative work~\cite{FuzzyLinkography}.

Closely related research on \emph{creative version control}~\cite{CreativeVersionControl} proposes four major themes that emerge from studying the use of version control in creative practice, two of which are of particular interest to us: ``using versions as a palette of materials'' and ``reflecting on and reusing versions across long time scales''. Creative version control systems like Quickpose~\cite{Quickpose}, Spellburst~\cite{Spellburst}, Sharp~\cite{Sharp}, and HistoryPalette~\cite{HistoryPalette} provide affordances for capturing, navigating, and reflecting on version histories of creative projects, often explicitly materializing past artifact versions or design moves in a way that permits the flexible reincorporation of ideas from prior design explorations into the ongoing design process.

Some recent AI-based creativity support tools, particularly image-editing canvases, combine provenance representations with finer-grained interaction with generated content. Expandora~\cite{Expandora}, for example, organizes exploration as intent-annotated branches, foregrounding relationships among alternative image-level directions; 
Other systems move closer to element-level reuse. ImaginationVellum~\cite{ImaginationVellum} materializes user-defined intent tags that guide nearby generations alongside provenance visualization of generated assets. Atelier~\cite{Protosampling} supports the extraction and recombination of visual assets while exposing their lineage across generated outcomes, and DesignTrace~\cite{DesignTrace} represents abstract design attributes that can be extracted and remixed across generated designs.


Across these systems, provenance remains primarily organized around generated assets, branches, or design states, even when finer-grained assets, intent tags, or attributes can be reused across generations. Trinketry instead makes designer-preserved elements themselves persistent units of the exploration history. A trinket therefore functions both as generation input and a persistent access point into the evolving trace. Trinketry thus shifts exploration history from image-level branches toward the elements designers deliberately choose to carry forward.





\subsection{Reflection on Creative Histories}
Much HCI work on reflective practice builds on \citeauthor{ReflectivePractitioner}~\cite{ReflectivePractitioner}, who distinguishes between ``reflection-in-action'' (the form of reflection that occurs during practice, for instance while a designer is designing an artifact) and ``reflection-on-action'' (the form of reflection that occurs retrospectively, when reviewing the outcomes of practice). Systems designed explicitly to promote reflection are often cast predominantly as scaffolding post-hoc sensemaking, for instance through visualization~\cite{ReflectionSpace}, documentation~\cite{Kaleidoscope,ReflectiveDesignDocumentation,ClearFairy}, or capture and replay~\cite{WatchMeWrite,Spin} of the design process. Some systems provide \emph{feedback}~\cite{proxona,Phraselette,UIFeedback,DisciplinedDrawing,SmithIDNCritique} or provocative \emph{questions}~\cite{IntentElicitation} that scaffold the elaboration of design intent in the moment, including by surfacing example artifacts that may help the user orient themselves in a broader design space~\cite{CorpusStudio,Shown,UIClip}; other systems attempt to help the user extrapolate or generalize a design space from concrete artifacts of interest~\cite{luminate,Schemex,Patchview,shen2025ideationweb}. Finally, some systems may support \emph{reflective revisitation} of past choices~\cite{ReflectiveCreators}. 
Most of those works scaffold reflection around whole artifacts, recorded states, or explicit design decisions. Trinketry draws inspiration from several of these approaches, but focuses predominantly on supporting retrospective sensemaking through visualization of element-level provenance, as well as in-the-moment reflection through a combination of example surfacing to facilitate revisitation (i.e., semantic trinket retrieval) and small-scale refinement of intent (i.e., affordances for selecting trinkets to retain).

\section{FORMATIVE STUDY}
\label{sec:formativestudy}
We conducted a formative interview study to understand how concept designers preserve, revisit, and trace promising elements during AI-assisted visual exploration, and what challenges arise in doing so.
We recruited four experienced professional concept designers (P1–P4; 4–15 years of experience, mean = 8), all with over two years of AI-assisted visual exploration experience and approximately one year of prior use of GenTune~\cite{GenTune}, an image-generation prototype with branching design history.
Participants were recruited through local art-community referrals and compensated approximately \$30 USD (Table~\ref{tab:demographics}).

We conducted 90-minute semi-structured interviews to examine participants’ AI-assisted visual exploration workflows and challenges. Each interview included two reflective design-tracing activities: one based on a user-selected GenTune session, and one based on a comparable session using commercial tools such as ChatGPT, Gemini, or Midjourney. Participants reconstructed their exploration process and discussed challenges in revisiting, tracing, and recovering prior design material. Interviews were audio-recorded, transcribed, and analyzed using reflexive thematic analysis~\cite{braun2019reflecting}. Initial coding was led by an author with prior professional experience as a concept designer. The second author reviewed the codes, and the two authors iteratively discussed and refined the coding scheme and themes.

\subsection{Findings}
\subsubsection{Designers Reuse Fragments, Not Whole Images}
Consistent with prior work~\cite{AIdeation, shi2025brickify, DesignPrompt2024}, participants rarely treated generated images as desirable wholes. Instead, they carried forward particular compositions, spatial structures, objects, lighting treatments, or atmospheres across iterations. As P2 explained, “\textit{I would first find the viewpoint I wanted, then try to combine [...] object references from earlier generations to see what result I could get.}” 

\subsubsection{Designers Revisit Meaningful Earlier States}
Participants reviewed earlier results to both inspect their design history and recover meaningful prior states for continued exploration. 
As P4 explained, “\textit{If I could trace back through the history step by step, I would better understand how the AI interpreted what it was told, and whether phrasing it differently might produce what I actually wanted}.”
Yet growing histories made this cumbersome; as P2 noted, “\textit{When there are too many generated images, I get too lazy to scroll back through them, and I would rather just start generating again.}”
These accounts motivate access to prior work through meaningful elements rather than chronological browsing alone.

\subsubsection{Drift Makes Exploration Continuity Difficult to Follow}
Participants described difficulty maintaining continuity when later generations unexpectedly changed desired structures or elements.
As P3 explained, “\textit{I wanted to see the image evolve step by step, but often the next generation would suddenly change the entire angle or composition}”. Such shifts were also difficult to interpret
: “\textit{I’m not really sure what exactly the image changed in response to}” (P4). 
Repeated attempts to restore desired features could further blur the original direction, with P1 noting that “\textit{after trying for a while, the exploration started to lose focus}.” As histories accumulated, drift made it increasingly difficult to trace where desired elements originated, how they changed, and whether they later resurfaced.


\subsection{Design Goals}
Across these findings, participants needed to preserve promising elements, trace how they changed across generations, and revisit prior material to recover alternatives and reflect on shifts in design intent. We therefore derived three design goals:

\begin{itemize}
    \item \textbf{DG1: Support preservation and recombination of promising elements across iterations.}
    Designers should be able to carry selected visual and semantic elements forward independently of their source images.

    \item \textbf{DG2: Support tracing how promising elements evolve across exploration.}
    Designers should be able to understand where important elements originated and how they were preserved, changed, lost, or reintroduced across iterations.

    \item \textbf{DG3: Support recovering relevant prior material for comparison and continued exploration.}
    Designers should be able to return to useful earlier elements and outcomes without exhaustively browsing an accumulating chronological history.
\end{itemize}


    


\begin{figure*}
    \centering
    \includegraphics[width=0.97\textwidth]{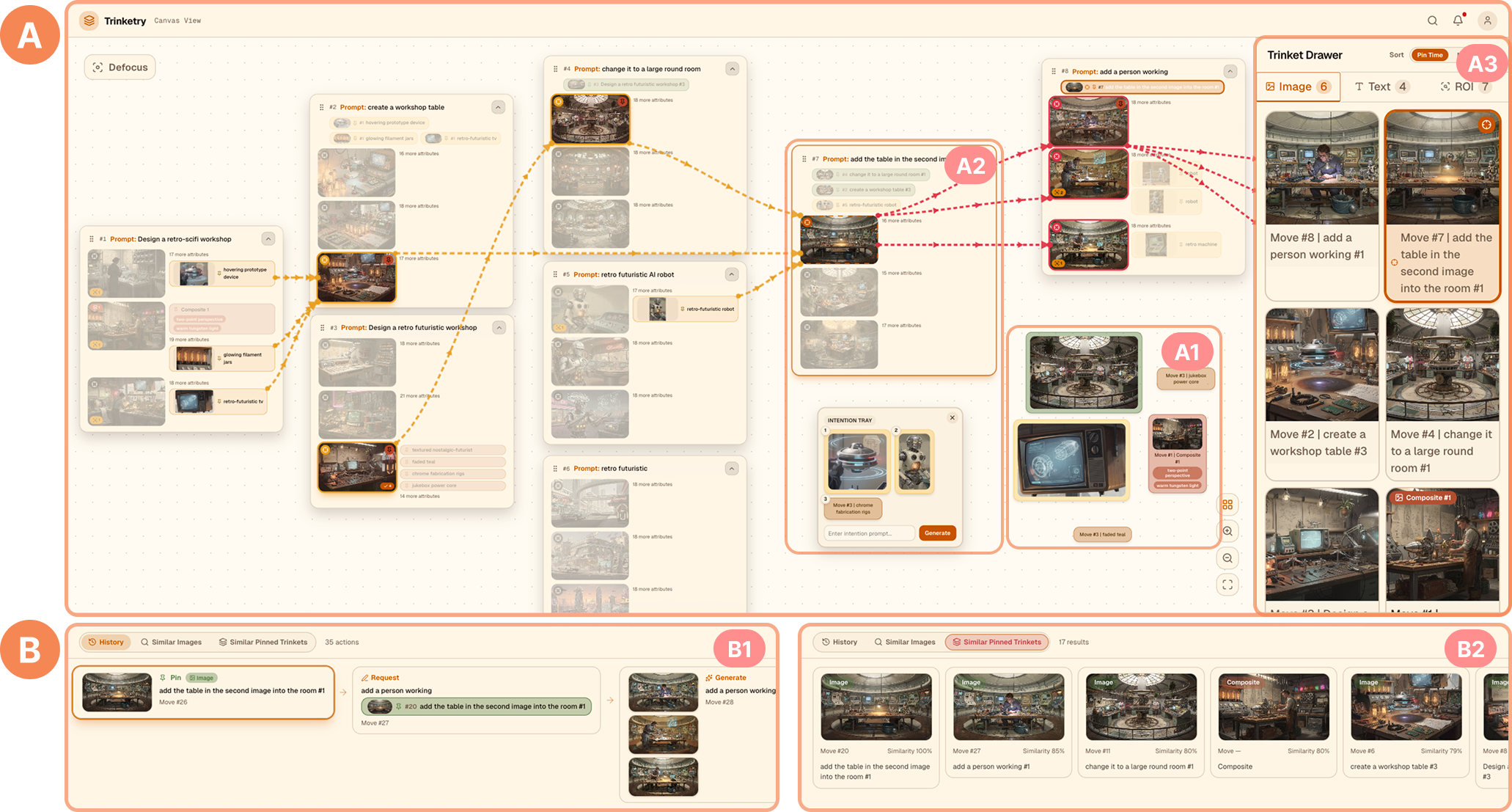}
    \caption{The Trinketry user interface with a trinket highlighted. The main canvas (A) serves as the workspace where trinkets (A1) and intention trays (A2) can be arranged freely. The provenance graph is rendered using yellow arrows for parent lineages and red arrows for child lineages. When a trinket is pinned, it is saved to the trinket drawer (A3), where trinkets are organized by category. The switchable bottom panel (B) supports both navigation and retrieval through the design history view (B1), which presents a left-to-right timeline of request, generation, pin, and unpin moves, and through the similar images and similar pinned trinkets views (B2), which rank prior artifacts by similarity to the currently highlighted trinket.}
    \label{fig:UserInferface}
    \Description[Trinketry interface showing canvas provenance, trinket drawer, and bottom history and retrieval panels.]{Annotated screenshot of the Trinketry user interface, divided into a large upper canvas region labeled A and a lower switchable panel labeled B. The canvas contains several intention trays arranged across a workspace, each showing prompts and generated retro-sci-fi workshop images. The highlighted trinket is the first image trinket inside the expanded intention tray in subregion A2, near the center-right of the canvas. Related items remain emphasized while unrelated trays and trinkets are faded. Dashed yellow arrows connect the highlighted trinket to earlier source images, and dashed red arrows connect it to later reused results, showing upstream parent lineage and downstream child lineage.
    
    Within A, subregion A1 shows reusable trinkets placed freely on the canvas outside trays: a larger image trinket with a green border, a larger ROI trinket with a yellow border, a smaller dark red-orange composite trinket with its image and labels, and two small light brown text trinkets. Subregion A2 shows two intention trays: one expanded tray with prompt text, input trinkets, and generated results, and one smaller tray with two ROI input trinkets, one text trinket, a prompt field, and a Generate button, illustrating recombination for new image generation. Subregion A3 is the Trinket Drawer on the right, organized by category tabs for image, text, and ROI trinkets.
    
    In B, panel B1 shows a left-to-right history timeline of moves such as pin, request, and generate, while B2 shows prior pinned trinkets ranked by similarity to the highlighted trinket. Together, the figure shows freeform arrangement, provenance tracing, history-based navigation, and semantic retrieval.}
\end{figure*}
\section{TRINKETRY}
\label{sec:system}
We instantiate our concept of user-preserved design elements in \emph{Trinketry}, a canvas-based system for AI-assisted visual exploration. Trinketry materializes elements that designers deliberately preserve as \emph{trinkets}: persistent objects that can be carried across generations and used to structure access to prior exploration.


\subsection{Interacting with Trinkets}
Trinkets support three interconnected forms of interaction aligned with our design goals. First, designers can preserve promising visual or semantic elements as reusable materials and recombine them in subsequent generations. Second, the same trinkets serve as trace anchors for navigating where elements originated and how they were reused or transformed. Third, trinkets serve as multi-modal retrieval queries for surfacing semantically related material from elsewhere in the exploration history. We describe these interactions below.

\subsubsection{Trinkets as Persistent Design Elements}
Trinketry distinguishes between generated artifacts and \emph{trinkets}. Generated images, regions, and semantic labels become trinkets only when a designer deliberately preserves them as elements worth carrying forward.

Because designers may preserve ideas at different levels of visual and semantic granularity, Trinketry supports four trinket representations:
\begin{itemize}
    \item \textbf{Image trinkets} preserve a complete generated image together with its associated text labels set as a reusable visual reference.

    \item \textbf{Region-of-interest (ROI) trinkets} preserve a designer-selected region of a generated image together with a revisable semantic label, allowing specific visual elements to be isolated from the surrounding output.

    \item \textbf{Text trinkets} preserve semantic concepts, either selected from system-suggested labels or entered manually by the designer.

    \item \textbf{Composite trinkets} preserve a generated image together with designer-selected labels, representing qualities that may depend on both visual and semantic information, such as composition, atmosphere, or viewpoint.
\end{itemize}

\subsubsection{Trinkets as Reusable Materials}
Designers preserve trinkets by pinning them to the \textit{Trinket Drawer} (Fig.~\ref{fig:UserInferface}-A3), keeping selected elements accessible independently of the generated images in which they appeared. Pinned trinkets can later be dragged back onto the canvas, freely arranged (Fig.~\ref{fig:UserInferface}-A1), and reused in new exploration directions.

To recombine trinkets, designers place them into \textit{intention trays} (Fig.~\ref{fig:UserInferface}-A2), canvas-based containers that group selected trinkets under a shared generation goal. 
Designers can gather visual and semantic trinkets from across the exploration, specify an instruction, and generate a batch of three images conditioned on their combination. After generation, the tray displays the prompt, input trinkets, and resulting images.
Newly extracted trinkets remain associated with their source tray and image while also being available for reuse elsewhere on the canvas.


\subsubsection{Trinkets as Trace Anchors}
A preserved trinket can serve as an entry point into the exploration history. When a designer highlights a trinket, Trinketry activates a trace view that reconstructs its upstream and downstream provenance relationships across the canvas (Fig.~\ref{fig:UserInferface}-A). The resulting graph shows where the trinket originated, which prior elements contributed to it, and which later generations reused or transformed it. Designers can highlight another connected trinket to recenter the trace around that element, allowing them to move through the history by following the trajectories of particular design ideas.

Trinketry complements this element-centered view with a chronological interaction history in the bottom panel (Fig.~\ref{fig:UserInferface}-B1), recording actions such as preserving trinkets and generating from intention trays. Each entry links back to the corresponding location on the canvas. Together, the history view shows when interactions occurred, while trinket-centered traces reveal how selected elements persisted, changed across generations, and contributed to later outcomes.

\subsubsection{Trinkets as Retrieval Queries}
A preserved trinket can also serve as a retrieval query for semantically related material elsewhere in the exploration history. When a designer highlights a trinket, Trinketry compares its available visual and textual information against previously generated images and preserved trinkets, surfacing the most similar results in dedicated panels in the bottom view (Fig.~\ref{fig:UserInferface}-B2). This multimodal retrieval can surface related material from otherwise disconnected branches, helping designers revisit overlooked alternatives, compare related elements, and reconnect with temporally non-
adjacent parts of the exploration.

\begin{figure*}
    \centering
    \includegraphics[width=0.97\linewidth]{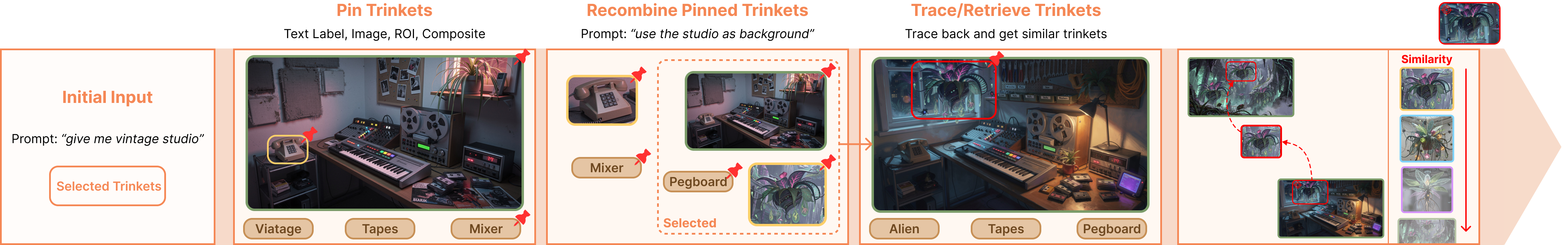}
    \caption{An example workflow in Trinketry. The designer extracts and pins trinkets from a generated result, then recombines a pinned image trinket with previously pinned ROI and text trinkets to generate a new output. They then select the new image to trace its provenance, highlight the alien plant element, and retrieve similar ROI trinkets.
    }
    \Description[Workflow showing pinning trinkets, recombining them in a new prompt, then tracing and retrieving related trinkets.]{Left-to-right workflow diagram illustrating element-centered exploration in Trinketry. The first panel, “Initial Input,” shows a prompt reading “give me vintage studio” and an empty selected-trinkets area. The second panel, “Pin Trinkets,” shows a generated studio scene from which several trinkets are extracted and pinned: the full image as an image trinket, a cropped telephone ROI trinket, a smaller crop from the upper-right of the image, and three text trinkets labeled “Viatage,” “Tapes,” and “Mixer.” The panel title indicates the supported pinned types are text label, image, ROI, and composite.
    
    The third panel, “Recombine Pinned Trinkets,” shows a new prompt, “use the studio as background.” Inside a dashed selected area are three visual trinkets: the pinned studio image, a pinned alien-plant ROI, and other reused trinkets including the telephone ROI and text trinkets such as “Mixer” and “Pegboard.” This illustrates recombining previously pinned image, ROI, and text trinkets in a new intention tray to generate another output.
    
    The fourth panel, “Trace/Retrieve Trinkets,” shows the resulting studio image with an alien plant inserted. Red boxes and dashed arrows trace the highlighted alien-plant element back to earlier source trinkets and forward to the new output. On the far right, a vertical similarity list ranks related ROI trinkets, with the most similar alien-plant crop at the top. Overall, the figure shows pinning extracted trinkets, recombining them to generate a new image, then tracing provenance and retrieving semantically similar trinkets.}
    \label{fig:workflow}
\end{figure*}

Fig. \ref{fig:workflow} illustrates how these roles connect across a single exploration workflow. A designer first preserves useful trinkets, later recombines them to generate a new result, and then uses a resulting element as an anchor for tracing its provenance and retrieving related prior material.

\subsection{Technical Implementation}
Trinketry is implemented as a web application. The frontend is built using Next.js, and the backend is built in Python using FastAPI for APIs and orchestration. GPT-5 mini~\cite{gpt-5-mini}
is chosen as the LLM, and FLUX.2 [pro]~\cite{flux-2-pro} 
is used for image synthesis. 

\begin{figure*}
    \centering
    \includegraphics[width=0.97\linewidth]{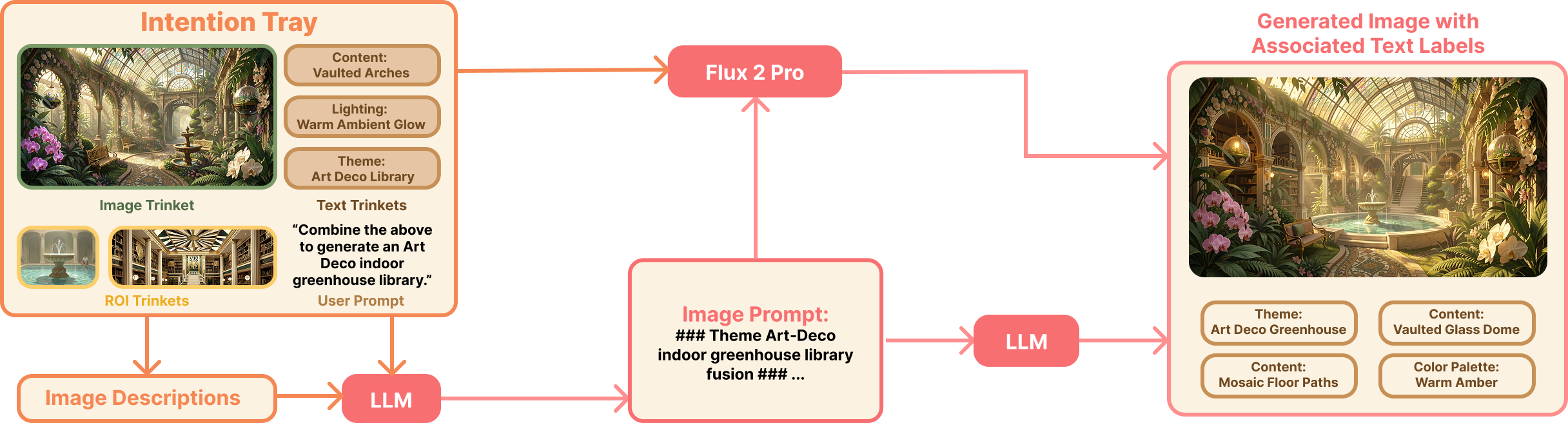}
    \caption{Image generation pipeline: 
    From an intention tray and user prompt, Trinketry uses one LLM to produce a refined image prompt, sends it with the source images of input image-type trinkets to FLUX.2 [pro] for image generation, and uses a second LLM to extract categorized text labels. 
    }
    \Description[Image generation pipeline from intention-tray trinkets and prompt to a generated image with categorized text labels.]{Left-to-right pipeline diagram showing image generation in Trinketry. On the left, an Intention Tray contains an image trinket showing an indoor greenhouse, two ROI trinkets depicting a fountain and an Art Deco library, three text trinkets labeled Content: Vaulted Arches, Lighting: Warm Ambient Glow, and Theme: Art Deco Library, and a user prompt reading, “Combine the above to generate an Art Deco indoor greenhouse library.”

    Image descriptions derived from the input image-type trinkets and the user prompt are passed to a first LLM, which produces a refined Image Prompt, illustrated as “Theme Art-Deco indoor greenhouse library fusion.” The refined prompt is sent to Flux 2 Pro together with the source images of the input image-type trinkets. Flux 2 Pro then generates a new image showing an Art Deco indoor greenhouse.

    The same refined image prompt is also passed to a second LLM, which extracts categorized text labels. Four example labels are shown: Theme: Art Deco Greenhouse, Content: Vaulted Glass Dome, Content: Mosaic Floor Paths, and Color Palette: Warm Amber. The generated image and these extracted labels are associated together as the pipeline output. Overall, the figure shows how Trinketry transforms an intention tray and user prompt into a refined prompt, generated image, and categorized text labels.}
    \label{fig:ImageGeneration}
\end{figure*}

\begin{figure*}
    \centering
    \includegraphics[width=0.97\linewidth]{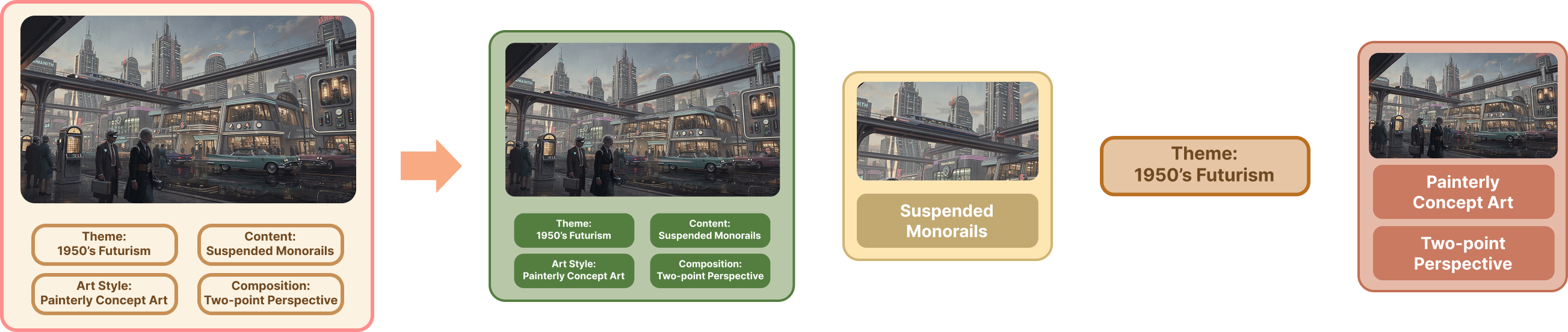}
    \caption{Four types of trinkets can be extracted from generated images. Image trinkets preserve the whole image and its associated labels. ROI trinkets preserve a cropped region, with a label suggested by matching its SigLIP2 embedding against candidate labels derived from the source image’s theme and content keywords. Text trinkets preserve individual labels. Composite trinkets pair an image with a designer-selected subset of its associated labels.
    }
    \Description[Generated image and labels represented as image, ROI, text, and composite trinkets.]{Left-to-right diagram showing four trinket representations derived from the same generated image and its semantic labels. The source is a retro-futuristic city scene with elevated monorails, paired with four labels: Theme: 1950’s Futurism, Content: Suspended Monorails, Art Style: Painterly Concept Art, and Composition: Two-point Perspective. The image trinket retains the complete image together with all four labels. The ROI trinket contains a cropped portion of the image showing the suspended monorails, paired with the label “Suspended Monorails.” The text trinket contains only the label “Theme: 1950’s Futurism.” The composite trinket retains the complete image but pairs it with only two selected labels, “Painterly Concept Art” and “Two-point Perspective.” The examples visually distinguish whether a trinket preserves the whole image and all labels, a selected visual region, an individual semantic label, or the whole image with a selected subset of labels.
    }
    \label{fig:TrinketConstruction}
\end{figure*}
\subsubsection{Image Generation and Trinket Extraction}
For image generation, Trinketry adapts GenTune's traceable prompting pipeline~\cite{GenTune} for multimodal generation. Each generated image is associated with semantic labels produced during prompt processing (Fig.~\ref{fig:ImageGeneration}).
Designers can extract trinkets from each generated image (Fig.~\ref{fig:TrinketConstruction}), and for ROI trinkets, designers may crop regions and revise the ROI labels suggested through similarity matching in SigLIP2’s embedding space~\cite{tschannen2025siglip}.

\subsubsection{Provenance Tracing}
Trinketry records provenance relationships among generated artifacts, preserved trinkets, and intention trays. Each intention tray stores the trinkets used for a generation request, and each resulting image is linked to that source tray. ROI and composite trinkets retain references to the source images from which they were preserved, while image trinkets take the mixed-type input trinkets recorded in the source tray as the parents. When a trinket is highlighted, Trinketry recursively follows these relationships upstream to its sources and downstream to later generations that reuse it.

\subsubsection{Multimodal Retrieval Across Trinket Representations}
Trinketry indexes preserved trinkets and generated images using their available visual and textual semantics, allowing a trinket to retrieve related material that is not explicitly connected through provenance. 

To compare two indexed items, we compute similarity scores across the available text--text, image--image, and cross-modal text--image channels using embedding sets encoded with SigLIP2~\cite{tschannen2025siglip}. Within each channel, we apply a symmetric mean-of-maxima pooling operator inspired by prior late-interaction scoring methods~\cite{colbert,faysse2025colpali,VideoColBERT} to capture mutual semantic correspondence while reducing sensitivity to unequal embedding-set sizes.

Because similarity scores scale differently across modality combinations due to the modality gap~\cite{liang2022mind}, we calibrate the available channel scores to a common range, then use the highest calibrated channel score as the similarity between two items. Appendix~\ref{app:retrieval-scoring} details the scoring formulation, calibration procedure, and evaluation.

\begin{figure}
    \centering
    \includegraphics[width=.95\linewidth]{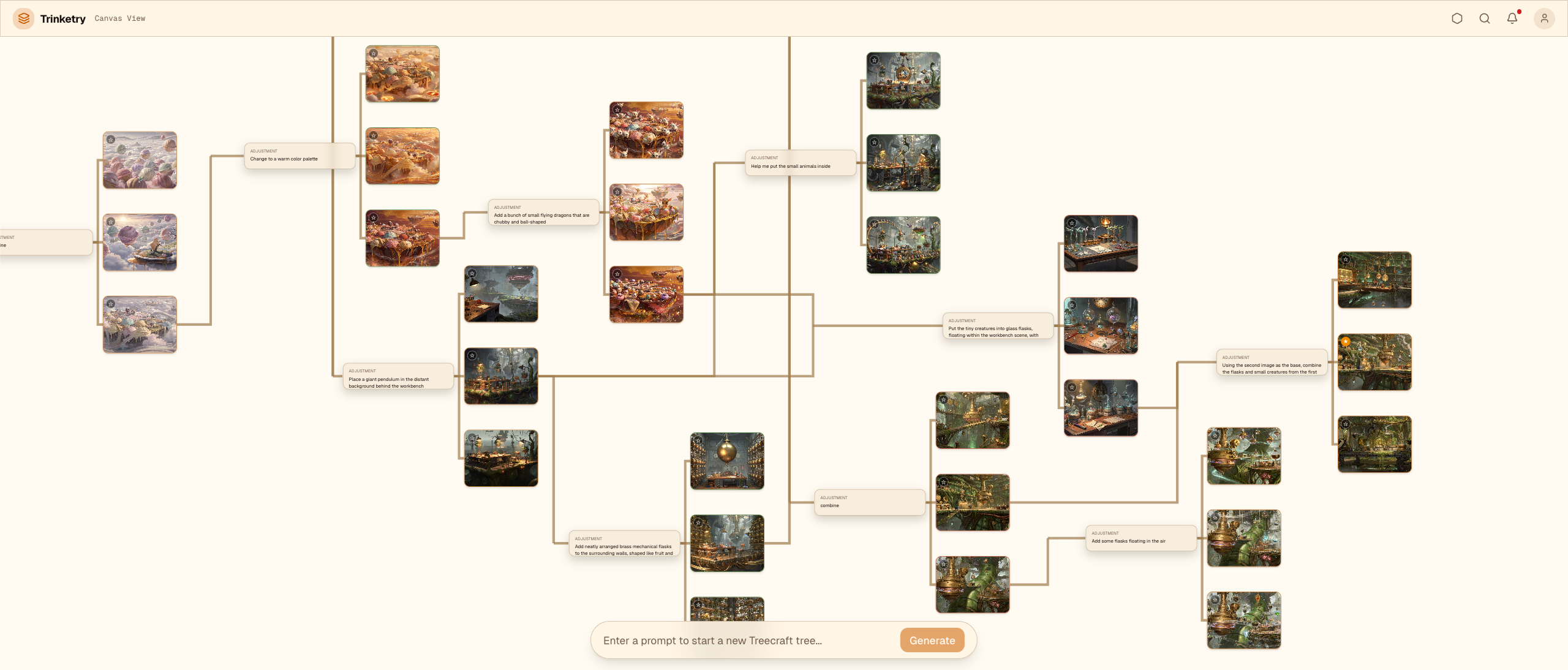}
    \caption{User interface of the baseline system, shown with P4's exploration result.}
    \Description[Baseline interface showing a draggable branching tree of prompts and generated image batches.]{Screenshot of the baseline system interface showing P4’s exploration result on a large beige canvas. Most of the screen is occupied by a left-to-right branching history structure. Each step in the exploration is shown as a small prompt or adjustment box connected by brown lines to a vertical stack of three generated image thumbnails. New branches extend from selected earlier results to later image stacks, making parent-child refinement relationships visible. In several places, connections from multiple earlier results feed into a later prompt box, indicating combination of more than one prior image.
    
    The image content evolves across the tree from pale floating landscape scenes on the left, to warmer orange fantasy scenes in the middle, and then to darker interior or glasshouse-like scenes with green tones on the right. The overall layout emphasizes chronological branching and revisitation of whole-image generations rather than reusable sub-elements. A thin top bar contains the application header and icons, while a prompt entry field with a Generate button sits at the bottom center for starting a new exploration branch. Overall, the figure illustrates the baseline system as a single-canvas interface organized around draggable branching history, where each node contains a prompt and its corresponding generated image batch.}
    \label{fig:baseline}
    \vspace{-10pt}
\end{figure}

\section{STUDY 1: ACTIVE VISUAL EXPLORATION}
\label{sec:study1}

Study 1 compared Trinketry with a simple branching history of generated images in an active visual exploration task with professional concept designers. We asked: \textbf{RQ1:} How do trinkets shape designers' active visual exploration and immediate access to prior design material compared with a whole-image branching baseline?

\subsection{Study Overview}
We used a within-subjects design comprising an active visual exploration task followed by a structured trace co-review of the preceding process.

\paragraph{Baseline}
To benchmark Trinketry, we developed a baseline matched on core generation capabilities (Fig. \ref{fig:baseline}). 
It supported the same core generation, refinement, and image-combination capabilities, with exploration history represented as a draggable prompt–image branching tree.
The baseline retained an explicit, navigable branching history but did not support trinkets: designers could reuse whole generated images, but could not preserve element-level units as persistent objects for reuse, tracing, or semantic retrieval.

We selected this representation because branching histories are common in recent AI-assisted visual exploration systems~\cite{GenTune, Expandora, DesignTrace, choi2026ideablocks}. This comparison allowed us to examine what changes when exploration history is organized around persistent designer-selected elements rather than primarily around whole-image branches.


\paragraph{Participants}
We recruited 12 professional concept designers with 3–21 years of experience (M=8.3; Appendix~\ref{tab:demographics}), including four participants from the formative study. Participants received approximately \$40.

\paragraph{Tasks and Procedure}

The study lasted approximately 2.5 hours and began with a 5-minute briefing. Participants completed one 30-minute environment concept-design task per condition, using either an \textit{Art Deco Research Room} or \textit{Retro-Futuristic Creative Workshop} brief; system order and design topics were counterbalanced across participants. Each condition included a 10-minute tutorial, the exploration task, an immediate trace co-review, and a post-condition questionnaire. Participants were encouraged to explore alternatives before selecting a final image.


During the trace co-review, participants revisited the preceding session with the experimenter using a structured interview protocol (see Appendix~\ref{Supplement_Trace_CoReview}). After both conditions, participants completed a comparative questionnaire and a 15--20-minute semi-structured interview reflecting on differences between the systems and the role of design histories in their exploration practice.

In line with a Comparative Structured Observation study design~\cite{mackay2025comparative}, we explicitly invited participants to reflect on the similarities and differences between their experiences using the two systems, and analyzed qualitative reflections in conjunction with quantitative data to more thoroughly characterize the impacts of each system's design.

\subsection{Measurements and Analysis}

\paragraph{Exploration behavior}
System logs captured generation and reuse activity throughout each session. We analyzed the amount and form of prior-material reuse across conditions and report generation counts to contextualize differences in exploration breadth and interaction pace.

\paragraph{Immediate trace review}
The structured co-review examined how participants used each history representation to revisit the process they had just completed. The interview covered the overall trajectory, major changes and turning points, meaningful elements, their origins and reuse, abandoned or overlooked material, and what participants might revisit if they continued the exploration. We analyzed these accounts to characterize how each representation supported immediate reconstruction of process structure, material history, and opportunities for continued exploration. These immediate post-task accounts also serve as the creator reference accounts for Study 2's delayed reconstruction analysis; we treat them as reference accounts rather than exhaustive ground truth.


\paragraph{Questionnaires}
After each condition, participants rated their experience reviewing the preceding process on 7-point Likert scales; conditions were compared using Wilcoxon signed-rank tests~\cite{wilcoxon}. After both conditions, participants rated their relative preference for Trinketry versus the baseline across reuse, traceability, retrieval, satisfaction, and perceived control on a 7-point scale (1 = baseline, 7 = Trinketry). We used one-sample Wilcoxon signed-rank tests against the neutral midpoint of 4~\cite{roberson1995analysis,capanu2006testing,taheri2013generalization}. 
We treat these ratings as descriptive, subjective evidence that contextualizes the behavioral logs and interview findings. We report Wilcoxon signed-rank statistics for transparency and completeness across all items.

\paragraph{Qualitative analysis}
All trace co-reviews and final interviews were audio-recorded, transcribed, and analyzed using reflexive thematic analysis~\cite{braun2019reflecting}. One author with prior professional concept-design experience led repeated transcript review and open coding around reuse, tracing, revisitation, and differences between conditions. Codes were iteratively grouped into higher-level themes through cross-participant and cross-condition comparison. Two additional authors discussed and refined the emerging themes with the lead coder, helping consolidate feature-level observations into broader patterns of exploration, control, and access to prior material.

\begin{figure}
    \centering
    \includegraphics[width=1\linewidth]{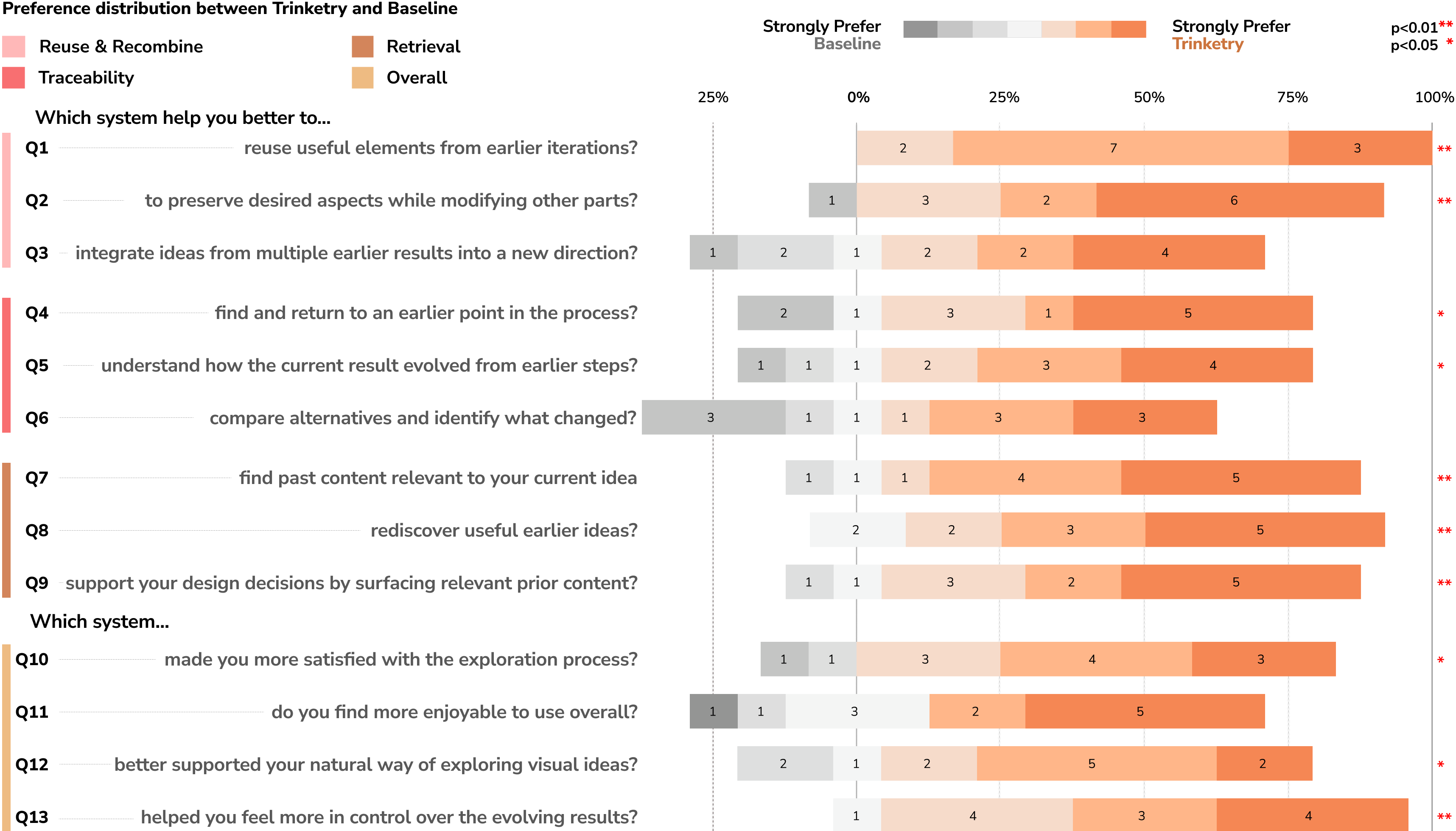}
    \caption{Results from the final preference questionnaire comparing Baseline and Trinketry across Reuse \& Recombine, Traceability, Retrieval and Overall Evaluation.}
    \Description[Diverging bar chart of final preferences, with most responses favoring Trinketry over the baseline.]{Diverging stacked bar chart of questionnaire preferences for Trinketry versus the baseline across 13 items. Each horizontal bar summarizes 12 participants on a 7-point preference scale, with gray segments to the left for stronger preference for the baseline and orange segments to the right for stronger preference for Trinketry. The questions are grouped into four sections: Reuse and Recombine: Q1, reuse useful elements from earlier iterations; Q2, preserve desired aspects while modifying other parts; Q3, integrate ideas from multiple earlier results into a new direction. Traceability: Q4, find and return to an earlier point in the process; Q5, understand how the current result evolved from earlier steps; Q6, compare alternatives and identify what changed. Retrieval: Q7, find past content relevant to your current idea; Q8, rediscover useful earlier ideas; Q9, support your design decisions by surfacing relevant prior content. Overall evaluation: Q10, made you more satisfied with the exploration process; Q11, find more enjoyable to use overall; Q12, better supported your natural way of exploring visual ideas; Q13, helped you feel more in control over the evolving results. Red significance markers indicate p less than 0.01 with two asterisks and p less than 0.05 with one asterisk.
    
    All 13 bars lean right, indicating an overall preference for Trinketry. The strongest result is Q1, where all 12 responses favor Trinketry. Strong preferences also appear for Q2, Q7, Q8, Q9, and Q13; these are marked significant, mostly at p less than 0.01. Q4 and Q5 also favor Trinketry, though less strongly. Q10 and Q12 show moderate but significant preference for Trinketry. The most mixed results are Q3, Q6, and Q11, which still lean toward Trinketry but sit closer to the midpoint and are not marked significant. No question shows a majority preference for the baseline. Overall, the chart indicates that participants preferred Trinketry most clearly for reuse, retrieval, and control, with weaker advantages for integrating ideas, comparing alternatives, and enjoyment.}
    \label{fig:result_within}
\end{figure}

\subsection{Findings}
We report three findings that characterize how trinkets shaped active visual exploration and immediate access to prior design material.

\subsubsection{Selected Elements Became Reusable Units of Exploration}
Trinketry changed the form of prior-material reuse from whole-image reuse toward reuse of designer-selected visual and semantic elements. Participants reused a similar number of prior-material inputs across conditions (154 in the baseline and 160 in Trinketry; Table~\ref{tab:reuse_logs}). 
In Trinketry, reuse was dominated by image and ROI trinkets, while text and composite trinkets were used more selectively. Participants' accounts suggest that image and ROI trinkets provided direct access to concrete visual material, whereas text and composite trinkets served more specialized roles for semantic concepts or qualities spanning visual and textual representations. 
Consistent with these behaviors, participants favored Trinketry in the direct comparative questionnaire for reusing useful material from earlier iterations and preserving desired aspects while modifying others (Fig.~\ref{fig:result_within}, Q1--Q2, $p < .01$; 100\% and 92\% favored Trinketry, respectively). 

Participants described this distinction as particularly useful when a generated image contained only partial value (P1--P3, P6--P9, P11). As P1 explained, ``\textit{Sometimes when I generate an image, I only want one element from it, and not the rest.}'' Trinkets allowed such fragments to be carried forward independently of their source images and recombined in later generations; as P5 described, ``\textit{I can select one part from A, one part from B, and turn them into something new.}''

\begin{table}[t]
\centering
\scriptsize
\setlength{\tabcolsep}{3pt}
\begin{tabular}{lcc}
\toprule
Measure & Base & Trinketry \\
\midrule
Reused elems.      & 154         & 160         \\
\quad Image        & 154 (100\%) & 69 (43.1\%) \\
\quad ROI          & 0           & 71 (44.4\%) \\
\quad Text         & 0           & 13 (8.1\%)  \\
\quad Comp.        & 0           & 7 (4.4\%)   \\

\bottomrule
\end{tabular}
\caption{Reuse behavior log summary.}
\Description[Table comparing reuse behavior logs for the baseline and Trinketry.]{Three-column table with measures in the first column and values for Base and Trinketry in the second and third columns. Total reused elements are similar across systems: 154 for the baseline and 160 for Trinketry. The main difference is the type of reused material. In the baseline, all reused elements are images: 154, or 100 percent. In Trinketry, reuse is distributed across multiple trinket types: 69 images, or 43.1 percent; 71 region-of-interest trinkets, or 44.4 percent; 13 text trinkets, or 8.1 percent; and 7 composite trinkets, or 4.4 percent.}
\label{tab:reuse_logs}
\vspace{-10pt}
\end{table}

\subsubsection{Element-Centered Traces Supported Focused Access to Prior Material}

\begin{figure}
    \centering
    \includegraphics[width=1\linewidth]{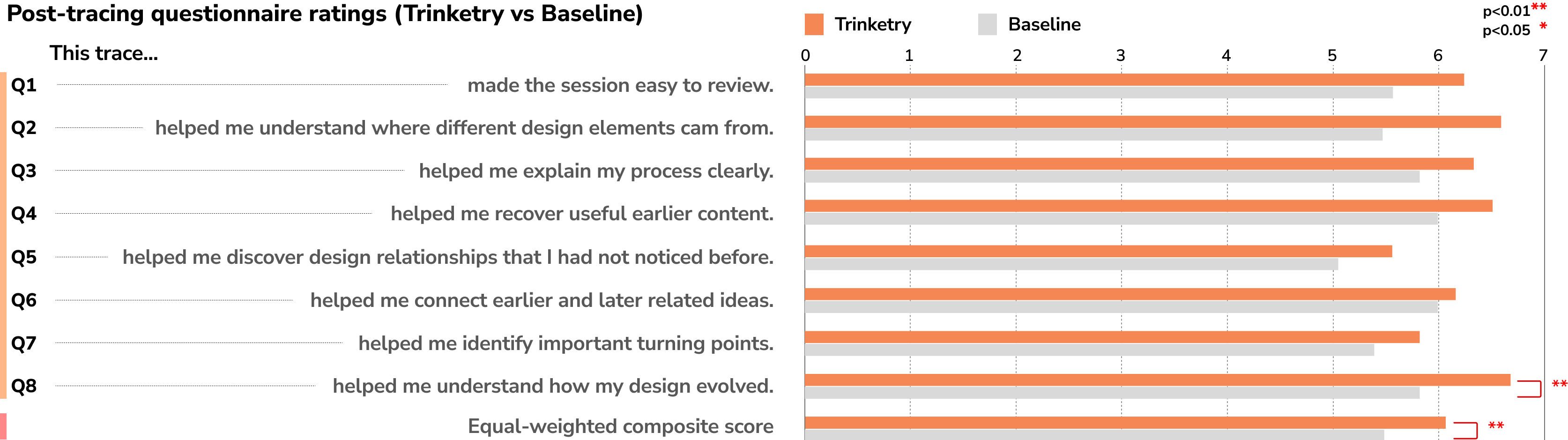}
    \caption{Results from the post-tracing questionnaire comparing Baseline and Trinketry on participants’ experience of reviewing and understanding their design process}
    \Description[Horizontal bar chart of post-tracing ratings, with Trinketry rated higher than the baseline on all items.]{Paired horizontal bar chart comparing post-tracing questionnaire ratings for Trinketry and the baseline on a 0 to 7 scale. Orange bars represent Trinketry and gray bars represent the baseline. The chart includes eight questionnaire items, Q1 through Q8, plus an equal-weighted composite score. The items ask whether the trace made the session easy to review, helped participants understand where different design elements came from, helped explain the process clearly, helped recover useful earlier content, helped discover unnoticed design relationships, helped connect earlier and later related ideas, helped identify important turning points, and helped understand how the design evolved. Red asterisks indicate significance thresholds of p less than 0.01 and p less than 0.05.
    
    For every item, the orange Trinketry bar is longer than the gray baseline bar, indicating consistently higher ratings for Trinketry. Most ratings for both systems fall between about 5 and 6.5, so both were viewed positively overall. The largest gap appears on Q8, “helped me understand how my design evolved,” and this item is marked significant at p less than 0.01. The equal-weighted composite score is also higher for Trinketry and marked significant at p less than 0.01. The remaining items show smaller advantages for Trinketry but are not marked significant in the chart. Overall, the figure shows that both systems supported retrospective review, but Trinketry was rated more favorably across all measures, with the clearest advantage in helping participants understand how their design evolved.}
    \label{fig:tracing_result}
\end{figure}

The two history representations supported different forms of access to prior work. The baseline exposed whole-image branches and their sequence directly, whereas Trinketry supported targeted inspection around preserved elements. 
As P3 noted, ``\textit{You can just follow the highlighted line and very quickly find the source.}'' Post-tracing ratings were high for both systems; only understanding how the design evolved differed significantly, favoring Trinketry (Fig.~\ref{fig:tracing_result}).

Trinketry also extended access beyond explicit lineage by retrieving semantically related prior materials from elsewhere in the exploration history. Participants found this useful when many alternatives had accumulated; as P5 explained, ``\textit{When you generate more than ten versions of the same image and later want to go back to find something similar from before, Trinketry becomes very useful.}'' 
Comparative ratings likewise favored Trinketry for understanding evolution and revisiting earlier states (Q4–Q5, $p<.05$), and for retrieving or rediscovering relevant prior material (Q7–Q9, $p<.01$); broad alternative comparison did not differ (Q6).

This focused access also introduced management costs. Participants (P2, P4, P8, P10) noted that as trinkets and generated results accumulated, the canvas became harder to organize and required additional navigation effort.


\subsubsection{Rapid Branching vs. Goal-Directed Exploration}
Participants described the two systems as supporting distinct exploration styles. This contrast was also reflected in the questionnaire: Trinketry was preferred as a more satisfying exploration process and a more natural way of exploring visual ideas (Fig.~\ref{fig:result_within}, Q10, Q12, $p < .05$), though not as more enjoyable (Q11).

Our findings suggest a tradeoff: the baseline supported faster, broader divergence across whole-image alternatives, whereas Trinketry supported deliberate and more goal-directed exploration.
Participants found the baseline more intuitive for rapid whole-image branching. As P8 noted, “\textit{I just pick two images and combine them directly.}” Its randomness also surfaced unexpected directions: P4 described it as producing “\textit{weird and unexpected things—it feels a bit like luck,}” which participants found useful when they “\textit{have nothing yet and want to look at a lot of possibilities}” (P5), especially when “\textit{the spec is still very vague}” (P2). This likely also explains why Trinketry was not significantly preferred for integrating ideas from multiple earlier results (Fig.~\ref{fig:result_within}, Q3), since the baseline more directly supported rapid whole-image merging.

Participants described Trinketry's additional selection effort made exploration slower but more deliberate.
As P3 explained, “\textit{You select things and pin them ... so it takes a bit more time to use.}” Participants felt this extra effort helped clarify what the design should carry forward. As P4 put it, it made clearer “\textit{what I want in the image—what elements should be there and how they should be presented.}”

This explicit selection strengthened participants’ sense of control. Participants preferred Trinketry for feeling more in control over evolving results (Fig.~\ref{fig:result_within}, Q13, $p < .01$). They first described this as more targeted control over outputs: “\textit{You can pull things out and generate [...] 
in a more targeted way, so it doesn’t feel like drawing a lottery. The first system feels more like gambling.}” (P2). Participants also described this as a form of anti-drift control, helping keep exploration aligned with task constraints and evolving intentions. As P1 noted, “\textit{Because I can collect elements and keep them on the side, they constantly remind me of my exploration goal.}”  In contrast, participants suggested that the baseline made it easier to lose focus; as P4 noted, it could “\textit{easily turn into a runaway horse.}” 

Together, trinkets supported more targeted reuse and retrieval and more deliberate, goal-directed exploration, at the cost of additional selection and management effort. Study 2 examines how these histories support sensemaking after a delay.


\section{STUDY 2: DELAYED SENSEMAKING AND CONTINUATION}

\subsection{Study Overview}

Study 2 was a longitudinal follow-up examining how generative design histories remain useful after designers become temporally distanced from their original exploration. Approximately four months after Study 1, returning participants revisited their prior Trinketry and baseline histories, continued their previous Trinketry exploration, and interpreted a Trinketry trace created by another designer. 

We investigate three complementary questions:

\begin{itemize}
    \item \textbf{RQ2:} How do trinket-centered and whole-image histories support designers in making sense of their own prior exploration after a delay?
    \item \textbf{RQ3:} How do preserved design elements support designers in resuming prior exploration after a delay?
     \item \textbf{RQ4:} How do external reviewers make sense of a designer's prior process from element-centered traces, and how does their sensemaking differ from the creator's delayed sensemaking of that same process?
\end{itemize}

\paragraph{Participants}
We recruited six returning participants (P1--P3, P6, P8, P12) from Study 1 via email. All had therefore previously completed both the Trinketry and baseline conditions and had first-hand experience with the design histories examined in this study. Participants received approximately US\$32 in compensation.

\subsection{Tasks and procedure}
The study lasted approximately two hours and consisted of three sequential parts corresponding to the three research questions: delayed self-sensemaking, continuation of prior work, and sensemaking of another designer's process.

\paragraph{Delayed self-sensemaking}
Participants first revisited the design histories emerging from their own work in both Study 1 conditions. For each history, they completed a 15-minute structured sensemaking interview using questions closely aligned with the Study 1 trace co-review (Appendix \ref{Supplement_Trace_CoReview}). We retained core questions about process structure, turning points, element provenance, and abandoned material, while additionally probing what remained memorable, uncertain, or recoverable after the delay. After reviewing both histories, participants completed a 15-minute comparative interview reflecting on differences between the two history representations. Following Comparative Structured Observation (CSO)~\cite{mackay2025comparative}, these paired reviews supported qualitative comparison of how the two representations shaped delayed sensemaking. The order of the Trinketry and baseline histories was counterbalanced across participants.

\paragraph{Continuation of prior work}
Second, participants returned to their previous Trinketry project and continued the exploration for approximately 25 minutes. As RQ3 focused specifically on how preserved design elements supported delayed continuation, this task used only the Trinketry history. Participants were free to revisit and reuse material from their existing history while developing the project further. This was followed by a 10-minute interview about how they used prior materials and traces in deciding how to proceed.

\paragraph{External sensemaking}
Finally, each participant spent approximately 15 minutes reviewing an anonymized Study 1 Trinketry trace produced by another Study 2 participant. Traces were assigned round-robin such that each was reviewed once and no participant reviewed their own work. Reviewers made sense of the process using the same analytic dimensions as in the self-sensemaking task and identified the trace evidence supporting their interpretations.

\subsection{Measurements and Analysis}

\paragraph{Data sources and analytic framework}
For the sensemaking analyses, we analyzed four complementary sources of data: Study 1 screen recordings, participants' immediate post-task Study 1 accounts, their delayed Study 2 sensemaking accounts, and the Study 2 external-review accounts. These sources served different epistemic roles. Screen recordings provided the behavioral record of observable actions during the original design sessions. The immediate Study 1 interviews instead provided a contemporaneous but subjective account of how creators understood their own process, including goals, evaluations, and decision rationales. 

We analyzed process accounts using a common framework derived from the structured questions used across both studies and refined through initial transcript coding. The framework captures three layers of the design process: \textit{process structure} (overall trajectory, major stages, convergence, and turning points), \textit{material history} (element salience, provenance, transformation and reuse, and abandoned or lost material), and \textit{design reasoning} (design goals and decision rationales). Definitions are provided in Table~\ref{tab:reconstruction_framework}. Across all analyses, we distinguished observable process evidence from inferred reasoning: trace structure, prompts, and provenance could support claims about actions, sequence, and material relationships, but were not treated as evidence of a designer's private rationale unless explicitly stated.

\paragraph{Claim extraction and framework coding}
One author with prior professional concept-design experience led the coding of the Study 1 reference accounts, Study 2 delayed sensemaking accounts, and Study 2 external-review accounts. Coding was carried out in two passes. First, the coder reviewed each transcript and extracted substantive claims about the design process while excluding procedural conversation and comments concerning only the interview or interface experience. Compound statements were decomposed into atomic claims, with each claim expressing a single proposition about an event, stage, element, goal, change, or rationale. Second, each claim was assigned one primary analytic dimension from the framework and, where substantively warranted, one secondary dimension.

\begin{table}[t]
\small
\renewcommand{\arraystretch}{1.08}
\setlength{\tabcolsep}{4pt}
\begin{tabular}{@{}p{0.25\columnwidth} p{0.68\columnwidth}@{}}
\toprule
\textbf{Code} & \textbf{Definition} \\
\midrule
\multicolumn{2}{@{}l}{\textbf{Alignment}} \\
Aligned & Study 2 and Study 1 claims express the same substantive meaning. \\
Partial & The claims overlap but differ in completeness, scope, or important detail. \\
Contradicted & The claims are substantively incompatible and cannot reasonably both be true. \\
Unverifiable & The Study 1 account provides insufficient evidence to assess the Study 2 claim. \\
\midrule
\multicolumn{2}{@{}l}{\textbf{Coverage}} \\
Recovered & The substantive meaning of a Study 1 claim is fully recovered in Study 2. \\
Partially recovered & Some, but not all, of this claim's substantive meaning is recovered in Study 2. \\
Not recovered & This claim's substantive meaning is not recovered in Study 2. \\
\bottomrule
\end{tabular}
\caption{Claim-level codes for alignment and coverage across sensemaking account.}
\Description[
Seven codes distinguish claim alignment from recovery of earlier claims across Study 1 and Study 2.
]{
Two-column table defining seven claim-level comparison codes in two groups: alignment and coverage. Alignment evaluates Study 2 claims against the corresponding Study 1 account. Its four codes are: Aligned, when the claims express the same substantive meaning; Partial, when they overlap but differ in completeness, scope, or important detail; Contradicted, when they are substantively incompatible; and Unverifiable, when the Study 1 account provides insufficient evidence to assess the Study 2 claim. Coverage reverses the comparison by evaluating how much of each Study 1 reference claim is recovered in Study 2. Its three codes are: Recovered, when the substantive meaning is fully recovered; Partially recovered, when only some of the substantive meaning is recovered; and Not recovered, when the substantive meaning is absent from the Study 2 account.
}
\label{tab:reconstruction_codes}
\end{table}

\paragraph{Alignment and coverage of process accounts}
We compared Study 2 accounts with the corresponding creator's immediate Study 1 reference account using two complementary claim-level analyses. \textit{Alignment} asks how closely a claim made during Study 2 corresponded to the closest comparable Study 1 claim. Each Study 2 claim was coded as \textit{Aligned}, \textit{Partial}, \textit{Contradicted}, or \textit{Unverifiable} according to the definitions in Table~\ref{tab:reconstruction_codes}. \textit{Coverage} reverses the comparison: for each Study 1 reference claim, we examined whether its substantive meaning was \textit{Recovered}, \textit{Partially Recovered}, or \textit{Not Recovered} in Study 2. Separating alignment from coverage distinguishes whether what participants said later corresponded to their earlier account from how much of that earlier account they recovered. 
Because claims were nested within participants, varied in number across conditions, and were sparse in some dimensions, we treat claim-level proportions as descriptive summaries rather than inferential estimates or effect-size comparisons. Because the Study 1 reference accounts capture only what participants articulated during the immediate interview, non-recovery in Study 2 indicates absence from the delayed account rather than direct evidence that the underlying memory was lost.

For claims concerning observable process structure or material history, we additionally consulted the Study 1 screen recording to resolve questions about actions, ordering, provenance, or reuse. For design reasoning, however, screen recordings could not establish private intentions or rationales. A rationale reported in Study 2 but absent from the creator's Study 1 account was therefore coded as \textit{Unverifiable}.

We applied this procedure to delayed creator sensemaking in both the Trinketry and baseline conditions and to external sensemaking of Trinketry traces. For the creator--reviewer comparison, the delayed creator and external reviewer accounts were independently compared against the same creator's Study 1 reference account.

\paragraph{Coding review and consensus}
A second author reviewed all extracted claims together with their source excerpts, framework assignments, cross-time claim matches, and alignment and coverage judgments. Ambiguous cases and disagreements were discussed until consensus was reached.

\paragraph{Comparative experience of the history representations}
The comparative interviews served two purposes: explaining patterns in the alignment and coverage analyses, and comparing participants' subjective experiences of the two history representations. 

Two authors independently coded the interviews for recurring similarities, differences, benefits, and limitations, then developed cross-participant themes through discussion. We triangulated these themes with the alignment and coverage results to relate participants' subjective assessments to the kinds and extent of process information they actually recovered.

\paragraph{Continuation Analysis}
For the continuation task, system logs and screen recordings were used to identify when participants returned to prior-session material and how that material entered subsequent generations. 
We distinguished reuse of trinkets already preserved during Study 1 from new trinkets extracted during Study 2 from earlier Study 1 generations, and recorded which continuation generations incorporated each form of prior material.
We paired these behavioral observations with participants' interview responses regarding how the prior trace shaped where they resumed exploration and where preserved material helped or constrained further development.


\begin{table}[t]
\small
\renewcommand{\arraystretch}{1.08}
\setlength{\tabcolsep}{4pt}
\begin{tabular}{@{}p{0.30\columnwidth} p{0.63\columnwidth}@{}}
\toprule
\textbf{Dimension} & \textbf{Definition} \\
\midrule

\multicolumn{2}{@{}l}{\textbf{Process structure}} \\
Overall process &
Overall trajectory of how the design developed across the session. \\

Main stages / branches &
Distinct phases, alternatives, or branches of exploration. \\

Convergence / taking shape &
When the design became recognizable or relatively stable. \\

Turning points &
Substantive changes in goal, approach, structure, style, or element role. \\

\midrule
\multicolumn{2}{@{}l}{\textbf{Material history}} \\
Element salience &
Which elements were treated as important or otherwise salient. \\

Element provenance &
Where an element originated and whether its continuity could be traced. \\

Transformation / reuse &
How elements were changed, reused, replaced, or recombined. \\

Lost / rejected / abandoned material &
Material that was rejected, abandoned, forgotten, or later recovered. \\

\midrule
\multicolumn{2}{@{}l}{\textbf{Design reasoning}} \\
Goal / intent &
What the designer was trying to achieve. \\

Decision rationale &
Why a change, selection, rejection, return, or other decision mattered. \\

\bottomrule
\end{tabular}
\caption{Analytic framework for making sense of prior design processes.}
\Description[
Ten analytic dimensions grouped into process structure, material history, and design reasoning.
]{
Two-column table defining ten dimensions for analyzing prior design processes, organized into three groups. Process structure contains four dimensions: overall process, describing the design's overall trajectory; main stages or branches, describing distinct phases, alternatives, or branches; convergence or taking shape, identifying when the design became recognizable or relatively stable; and turning points, capturing substantive changes in goal, approach, structure, style, or element role. Material history contains four dimensions: element salience, identifying important or otherwise salient elements; element provenance, describing where an element originated and whether its continuity could be traced; transformation or reuse, describing how elements were changed, reused, replaced, or recombined; and lost, rejected, or abandoned material, covering material that was rejected, abandoned, forgotten, or later recovered. Design reasoning contains two dimensions: goal or intent, describing what the designer was trying to achieve; and decision rationale, describing why a change, selection, rejection, return, or other decision mattered.
}
\label{tab:reconstruction_framework}
\end{table}


\subsection{Findings}

Our findings reveal four patterns in what different design-history representations made recoverable after a delay, how preserved material supported continuation, and where trace-based sensemaking remained limited.

\subsubsection{Preserved Elements Became Cues for Delayed and External Sensemaking}
At the claim level, delayed accounts of Trinketry traces showed descriptively stronger recovery of major stages and turning points in the design process than accounts of baseline branching image histories, while moment of convergence was broadly recoverable from both representations (Table \ref{tab:study2_reconstruction_results}). Participants' accounts suggest that preserved elements served not only as records of individually desirable materials, but also as landmarks for recognizing broader stages and transitions. P8, for example, reported that ``\textit{reviewing the trace of elements recovered a nearly forgotten memory of changing the entire spatial structure and exploration direction.}''

Preserved elements also provided entry points into local material histories, indicating which elements had mattered and where they came from. Element salience was more precisely recovered with Trinketry, while provenance was highly recoverable in both representations (Table \ref{tab:study2_reconstruction_results}). P6 described the trinket tray as an ``\textit{index directory}'' of previously selected material and explained that encountering a saved element ``\textit{reminded me that I had once explored this matter,}'' helping them recover the diamond as the central subject around which the researchers were organized. Trinketry's explicit source and reuse relations further supported tracing particular elements through the process. These cues also helped participants recover previously rejected or abandoned material, which was more frequently recovered with Trinketry than with the baseline (Table \ref{tab:study2_reconstruction_results}). P6, for example, explained that ``\textit{the line just now helped me find it back}'' before recovering a discarded alternative and recalling that it ``\textit{was not very attractive, so it was not placed.}''

The same cues were partially legible to external reviewers despite their lack of autobiographical memory. Reviewers recovered many major stages, convergence points, and provenance relations from Trinketry traces (Table \ref{tab:study2_reconstruction_results}). P8 described the focus function as particularly useful for identifying which elements the creator had used and retained. They later explained that reconstruction ``\textit{mainly still depends on tracking [an element], and then being able to see its source,}'' which allowed them to ``\textit{roughly trace its... context}'' and understand ``\textit{how it starts from one thing and then transforms into another thing.}'' Together, these results suggest that preserved elements functioned as trace anchors at two scales: locally, by exposing material histories, and more broadly, by providing landmarks for making sense of how the exploration developed.

\subsubsection{Element-Centered and Whole-Image Histories Supported Complementary Sensemaking}

The two history representations supported delayed sensemaking at different scales. Whole-image histories provided a clearer global chronological scaffold, while element-centered histories supported more precise inspection of particular materials and their relationships. Consistent with this perceived global advantage, baseline delayed accounts covered all tagged overall-process claims. Participants similarly described the whole-image tree as useful for seeing the session as a whole: P8 emphasized ``\textit{the whole context of the ideation,}'' P3 described easier to remember ``\textit{how I developed to the final step,}'' and P2 likened the representation to a concise ``\textit{report}'' or ``\textit{powerpoint}'' that exposed the overall trend. By foregrounding sequence and branching among complete states, the tree provided an immediately legible scaffold for narrating the overall trajectory.

This global coherence, however, did not always preserve the significance originally assigned to individual changes. Whole-image histories helped participants reconstruct what happened and in what order, but not necessarily how each change had been interpreted at the time. In the initial Baseline interview, P8 described a distinct branch involving “\textit{different compositions}” and “\textit{something like a split-level layout.}” During delayed sensemaking, however, P8 characterized the process as continued modification of one composition, noting that “\textit{the overall major direction hasn’t changed much.}” Thus, chronological visibility could make transitions easy to follow without necessarily preserving which of them had mattered as conceptual turns.

Element-centered histories instead foregrounded relationships around particular materials. P2 described Trinketry as more suitable for ``\textit{internal review}'' or detailed explanation because connected materials and their associated information were directly exposed. This relational precision, however, required participants to integrate multiple local histories into a larger chronology. P8 noted that once Trinketry's blocks were spatially rearranged, ``\textit{the context from nothing to something would be completely disrupted,}'' while P6 said reconstructing the overall story ``\textit{would take more time because elements could originate from different images}'' . Taken together, these complementary strengths suggest a representational tradeoff between maintaining a coherent global narrative and exposing detailed local histories.

\subsubsection{Preserved Elements Supported Delayed Continuation}
Trinketry's preserved exploration histories provided both entry points for resuming prior work and reusable materials for developing it further. Five participants reused old-session material in their very first continuation generation; the sixth did so by their second generation. P2 returned to ``\textit{something I regretted from the previous session},'' explaining that review helped them recover ``\textit{the sofa and television that had previously been abandoned, as well as the bookshelf},'' which they then brought back into the design. 
P12 similarly said the review ``\textit{helped me know where to at least start},'' leading them to revisit an unfinished part of the earlier story and prompted P12's subsequent exploration: ``\textit{After adding it, I thought I could try adding other elements and subjects.}''

Preserved material also remained actively useful throughout continuation. All six participants reused previously saved trinkets, which appeared in 26 of 53 continuation generations (Table~\ref{tab:continuation_inputs}). P8 explained that saved intermediate material let them ``\textit{try different variations}'' and ``\textit{pull it out directly and branch out from there}.'' Participants also rediscovered useful material that they had not preserved during Study 1: all six extracted new trinkets from old-session generations, and five subsequently used these selections across nine continuation generations. P6, for example, ``\textit{pinned some keywords and then used them to generate a new background},'' drawing on an earlier image's ``\textit{theme, style, and color palette}.'' 

However, this continuity might also constrain departure from earlier directions, potentially exacerbating certain varieties of design fixation~\cite{DesignFixation,DesignFixationGero,DesignFixationAI}. P8 noted that although they explored several variations, ``\textit{it was difficult to break out of that existing framework}'' because they kept reusing elements already present. P12 likewise questioned: ``\textit{if I were to develop a new image concept later, I am not sure I could use them directly, because they are so specific.}''
Preserved elements therefore made earlier directions immediately actionable after a delay, but the same persistence could also anchor subsequent exploration to the design space already established.


\begin{table*}[t]
\centering
\scriptsize
\setlength{\tabcolsep}{3.5pt}
\renewcommand{\arraystretch}{1.10}

\resizebox{\textwidth}{!}{
\begin{tabular}{@{}lccccccccc@{}}
\toprule
\textbf{Participant}
& \textbf{None}
& \textbf{P only}
& \textbf{R only}
& \textbf{N only}
& \textbf{P+R}
& \textbf{P+N}
& \textbf{R+N}
& \textbf{P+R+N}
& \textbf{Total} \\
\midrule

P1  & 3 & 0 & 0 & 4 & 2 & 3 & 1 & 1 & 14 \\
P2  & 0 & 1 & 2 & 2 & 0 & 5 & 0 & 0 & 10 \\
P3  & 1 & 1 & 0 & 1 & 0 & 1 & 0 & 1 & 5  \\
P6  & 0 & 1 & 0 & 3 & 1 & 3 & 0 & 0 & 8  \\
P8  & 1 & 2 & 0 & 4 & 0 & 1 & 1 & 0 & 9  \\
P12 & 4 & 1 & 0 & 0 & 0 & 2 & 0 & 0 & 7  \\

\midrule
\textbf{Total}
& \textbf{9}
& \textbf{6}
& \textbf{2}
& \textbf{14}
& \textbf{3}
& \textbf{15}
& \textbf{2}
& \textbf{2}
& \textbf{53} \\

\textbf{Mean $\pm$ SD}
& \textbf{1.50 $\pm$ 1.64}
& \textbf{1.00 $\pm$ 0.63}
& \textbf{0.33 $\pm$ 0.82}
& \textbf{2.33 $\pm$ 1.63}
& \textbf{0.50 $\pm$ 0.84}
& \textbf{2.50 $\pm$ 1.52}
& \textbf{0.33 $\pm$ 0.52}
& \textbf{0.33 $\pm$ 0.52}
& \textbf{8.83 $\pm$ 3.06} \\

\bottomrule
\end{tabular}
}

\caption{Composition of trinket inputs across continuation generations.
P denotes trinkets saved during the earlier session;
R denotes trinkets newly extracted during continuation from old-session generations; and
N denotes trinkets derived from continuation-session generations.
None indicates that no trinket was used as a generation input.}
\label{tab:continuation_inputs}
\Description[
Counts of continuation generations by combinations of earlier, recovered, and newly developed trinkets.
]{
Table showing the trinket-input composition of 53 continuation generations across six participants. Inputs are categorized as trinkets saved during the earlier session, trinkets newly extracted during continuation from old-session generations, trinkets derived from continuation-session generations, combinations of these sources, or no trinket input. The most common input composition combines previously saved and continuation-session trinkets, appearing in 15 generations, followed by continuation-session trinkets alone in 14 generations. Nine generations use no trinket input, six use only previously saved trinkets, and three combine previously saved trinkets with newly extracted old-session trinkets. Newly extracted old-session trinkets alone, their combination with continuation-session trinkets, and combinations of all three sources each occur in two generations. Across categories, previously saved trinkets appear in 26 of 53 generations, newly extracted old-session trinkets in 9, and continuation-session trinkets in 33. Participant totals range from 5 to 14 continuation generations.
}
\end{table*}

\subsubsection{Actions and Material Histories Were More Recoverable Than Design Rationales}

Across the analytic framework for sensemaking, observable process structure and material histories were generally more recoverable than the reasoning behind them (Table~\ref{tab:study2_reconstruction_results}).
Within design reasoning, broad goals remained relatively recoverable in both conditions, whereas the rationales behind particular decisions were substantially more fragile (Table~\ref{tab:study2_reconstruction_results}). Only 3 of 16 Trinketry rationales were fully recovered and none of the Baseline rationales were. Creators therefore often retained what they had been trying to achieve while losing or weakening why particular changes or selections had been made.

This limitation appeared in both representations because visible actions did not necessarily preserve their reasons. P6, after reviewing the baseline history, described wondering, ``\textit{Why did I think that way back then?}'' Even with Trinketry, they encountered a visible generation but were ``\textit{quite puzzled about why they had been prompting for images of this type}''. In some cases, Trinketry provided additional cues for recovering the reasoning associated with visible actions. P2 described: ``\textit{kind of like a note-taking function… I can easily recall what I was doing at that time.}'' 
Saved elements, prompts, and recombination relationships could therefore reactivate associated intentions without explicitly encoding them.

The distinction between action and reasoning became clearer in external review, where recovery of goals and rationales was lower than for delayed creators (Table~\ref{tab:study2_reconstruction_results}). Reviewers could often see that elements had been generated, preserved, combined, or rejected, but could not reliably determine why. Reviewing P12's trace, P8 reconstructed the generated components and their combination but not the creator's underlying story logic, noting, ``\textit{I can't see the initial context,}'' and questioned why some earlier elements were not used. The trace exposed the components and their combination, but not the narrative relationships that made those components meaningful to the creator.

Yet traces could occasionally preserve observable relations that creators themselves no longer mentioned. For P3, the external reviewer noticed that stained glass had failed to carry into one combination and was later ``\textit{called back}'' successfully, a lineage the delayed creator omitted; the Study 1 creator's account, however, additionally explained that colored glass was introduced because ``\textit{it would make the ceiling more interesting.}'' This qualifies the overall creator advantage: reviewers could occasionally recover trace-visible transformations or transitions that creators no longer mentioned, while still lacking the rationale that gave those actions meaning.

\begin{table*}[t]
\centering

\scriptsize
\setlength{\tabcolsep}{2.3pt}
\renewcommand{\arraystretch}{1.10}

\resizebox{\textwidth}{!}{
\begin{tabular}{@{}l
cccc
cccc
cccc@{}}
\toprule
& \multicolumn{4}{c}{\textbf{Trinketry Self}}
& \multicolumn{4}{c}{\textbf{Baseline Self}}
& \multicolumn{4}{c}{\textbf{Trinketry External}} \\
\cmidrule(lr){2-5}
\cmidrule(lr){6-9}
\cmidrule(l){10-13}

\textbf{Dimension}
& \textbf{SA} & \textbf{BA} & \textbf{FC} & \textbf{AC}
& \textbf{SA} & \textbf{BA} & \textbf{FC} & \textbf{AC}
& \textbf{SA} & \textbf{BA} & \textbf{FC} & \textbf{AC} \\
\midrule

\multicolumn{13}{@{}l}{\textit{Process structure}} \\

Overall process
& 15/19 (79\%) & 19/19 (100\%) & 1/8 (13\%) & 6/8 (75\%)
& 15/21 (71\%) & 18/21 (86\%) & 3/10 (30\%) & 10/10 (100\%)
& 6/8 (75\%) & 7/8 (88\%) & 2/8 (25\%) & 6/8 (75\%) \\

Main stages / branches
& 15/19 (79\%) & 18/19 (95\%) & 5/9 (56\%) & 9/9 (100\%)
& 11/18 (61\%) & 17/18 (94\%) & 3/10 (30\%) & 7/10 (70\%)
& 9/11 (82\%) & 10/11 (91\%) & 2/9 (22\%) & 7/9 (78\%) \\

Convergence / taking shape
& 6/6 (100\%) & 6/6 (100\%) & 6/8 (75\%) & 8/8 (100\%)
& 4/8 (50\%) & 8/8 (100\%) & 4/6 (67\%) & 6/6 (100\%)
& 4/4 (100\%) & 4/4 (100\%) & 4/8 (50\%) & 7/8 (88\%) \\

Turning points / direction changes
& 6/8 (75\%) & 7/8 (88\%) & 2/8 (25\%) & 7/8 (88\%)
& 2/8 (25\%) & 3/8 (38\%) & 1/9 (11\%) & 5/9 (56\%)
& 3/9 (33\%) & 7/9 (78\%) & 0/8 (0\%) & 4/8 (50\%) \\

\midrule
\multicolumn{13}{@{}l}{\textit{Material history}} \\

Element salience
& 8/11 (73\%) & 11/11 (100\%) & 4/10 (40\%) & 9/10 (90\%)
& 3/13 (23\%) & 10/13 (77\%) & 1/6 (17\%) & 5/6 (83\%)
& 5/12 (42\%) & 10/12 (83\%) & 2/10 (20\%) & 9/10 (90\%) \\

Element provenance
& 17/20 (85\%) & 18/20 (90\%) & 8/13 (62\%) & 13/13 (100\%)
& 18/20 (90\%) & 20/20 (100\%) & 10/14 (71\%) & 14/14 (100\%)
& 10/13 (77\%) & 11/13 (85\%) & 6/13 (46\%) & 11/13 (85\%) \\

Transformation / reuse
& 9/15 (60\%) & 13/15 (87\%) & 7/22 (32\%) & 16/22 (73\%)
& 10/15 (67\%) & 14/15 (93\%) & 4/24 (17\%) & 19/24 (79\%)
& 17/28 (61\%) & 25/28 (89\%) & 2/22 (9\%) & 17/22 (77\%) \\

Lost / rejected / abandoned
& 11/18 (61\%) & 15/18 (83\%) & 5/16 (31\%) & 13/16 (81\%)
& 4/9 (44\%) & 8/9 (89\%) & 1/12 (8\%) & 5/12 (42\%)
& 1/1 (100\%) & 1/1 (100\%) & 2/16 (13\%) & 7/16 (44\%) \\

\midrule
\multicolumn{13}{@{}l}{\textit{Design reasoning}} \\

Goal / intent
& 11/14 (79\%) & 13/14 (93\%) & 6/18 (33\%) & 16/18 (89\%)
& 8/16 (50\%) & 11/16 (69\%) & 1/17 (6\%) & 14/17 (82\%)
& 7/12 (58\%) & 10/12 (83\%) & 1/18 (6\%) & 11/18 (61\%) \\

Decision rationale
& 9/19 (47\%) & 15/19 (79\%) & 3/16 (19\%) & 12/16 (75\%)
& 5/15 (33\%) & 10/15 (67\%) & 0/13 (0\%) & 9/13 (69\%)
& 1/4 (25\%) & 3/4 (75\%) & 0/16 (0\%) & 10/16 (62\%) \\

\bottomrule

\end{tabular}
}
\caption{Descriptive claim-level alignment and coverage across analytic dimensions in Study 2. Claims are nested within participants, and participants contributed different numbers of claims across dimensions and conditions; percentages therefore summarize coded claims rather than participant-level rates and are not used for statistical inference. Strict alignment (SA) includes only Aligned Study 2 claims; broad alignment (BA) includes Aligned and Partial claims. Full coverage (FC) includes only Recovered Study 1 reference claims; any coverage (AC) includes both Recovered and Partially Recovered claims. Values are count/total (\%).}
\vspace{2pt}
\label{tab:study2_reconstruction_results}
\Description[
Study 2 alignment and coverage by analytic dimension for Trinketry self, baseline self, and external review.
]{
Table comparing claim alignment and coverage across ten analytic dimensions for Trinketry self-review, baseline self-review, and Trinketry external review. Results are grouped into process structure, material history, and design reasoning. For process structure, Trinketry self-review showed stronger recovery of main stages and turning points than baseline self-review, while convergence was broadly recoverable in all three comparisons. Baseline self-review showed the highest any-coverage rate for the overall process, reaching 100 percent compared with 75 percent for Trinketry self-review and external review. For material history, element salience was more precisely aligned in Trinketry self-review than baseline self-review, and element provenance was highly recoverable in both self-review conditions. Lost, rejected, or abandoned material showed substantially higher any coverage in Trinketry self-review, 81 percent, than in baseline self-review, 42 percent, or external review, 44 percent. Design reasoning was generally less recoverable than observable process and material history. Goal or intent remained relatively recoverable for creators, with any coverage of 89 percent in Trinketry and 82 percent in baseline, but decreased to 61 percent for external reviewers. Decision rationale was the weakest dimension: full coverage was 19 percent for Trinketry self-review and zero percent for both baseline self-review and external review. Overall, the table shows that observable stages, elements, and provenance were more recoverable than the reasons behind design decisions, while Trinketry and baseline histories provided different strengths for reconstructing the process.
}
\end{table*}

\section{DISCUSSION, FUTURE WORK, LIMITATIONS}
\label{sec:discussion}
\paragraph{From Traceable Actions to Shareable Rationale.}
Prior work such as DesignTrace shows how externalized attributes and histories can reify design rationale, helping designers articulate and revisit decisions during ongoing work~\cite{DesignTrace}. Our Study~1 similarly found that materialized elements supported designers in explaining and revisiting their process during immediate review (Fig \ref{fig:tracing_result}). Study~2, however, reveals an important temporal boundary: after four months, designers could often reconstruct what they had done, while the rationales behind particular decisions were substantially less recoverable. 
This suggests that traces which primarily externalize actions and materials may provide weaker support for recovering rationale over time unless decision context is also explicitly recorded.
This gap was even clearer for external reviewers, who could follow visible material relationships but often lacked the context needed to explain why they mattered. Traces intended for long-term reconstruction or collaboration may therefore need to preserve \emph{decision context} alongside provenance.


However, our Study~2 findings also suggest that trinket-centered action traces could sometimes cue designers' recollection of rationale through prior actions such as extraction, recombination, transformation, and repeated reuse. These action traces can provide concrete reminders of decision rationale. This complements prior approaches that scaffold rationale articulation through reflective documentation~\cite{Kaleidoscope} or in-situ questioning that prompts creators to explain decisions during the creative process~\cite{ClearFairy,IntentElicitation}. Our findings suggest a complementary opportunity, that the action trace itself could provide a scaffolding for such elicitation. Actions such as selecting, transforming, recombining, or abandoning material identify meaningful moments to which designers could attach lightweight explanations during or shortly after exploration. Traces could thus serve as anchors for capturing why an element mattered, why a direction changed, or why an alternative was rejected, preserving decision context alongside the actions and materials already recorded.


Embedding such rationale could also make design traces more useful as communicative artifacts. Within teams, traces could support asynchronous handoff by conveying both how design elements evolved and why particular directions were pursued. For clients or directors, they could expose explored alternatives and the reasoning behind a proposed direction. Traceable histories may also help distinguish human and AI contributions in complex workflows~\cite{HaLLMark,DraftMarks} and mitigate the ``AI memory gap'' in recalling one's own contributions~\cite{AIMemoryGap}. Future work could investigate how rationale-enriched design histories support handoff, collaboration, and review.

\paragraph{Rapid Iteration vs. Slow Deliberation in Generative Design}
Generative AI makes it easy to produce many alternatives, but cheaper generation does not necessarily lead to better exploration. Prior work notes that text-to-image prompting can devolve into “brute-force trial and error” that feels random and unprincipled~\cite{DesignGuidelines}, and that rapid prompting and random generation can increase the number of alternatives without adequately supporting their organization or reuse~\cite{HistoryPalette}. 
Under such conditions, we find designers slip into a 
“gambling” dynamic, in which generating again becomes easier than making sense of what has already been produced. While randomness can support broad exploration~\cite{AIdeation}, it can also contribute to intention drift and the accumulation of weakly organized results. In this sense, generative variability is not inherently beneficial; it is something designers need help managing~\cite{DesignPrinciplesGenAI}. 
Trinketry can here be understood as introducing productive friction: it slows users down in ways that pay off by supporting more deliberate choices about element selection, reuse, composition, and retrieval, thereby improving control over subsequent iterations. This aligns with prior HCI work suggesting that friction can be valuable when it interrupts mindless interaction and creates opportunities to reflect~\cite{AI_Instruments,SlowTechnology,DesignFrictions}. 

\paragraph{Decision Granularity in Co-Creativity}
Trinketry's element-centric interaction paradigm may be generalized to a wide variety of creative domains in which generative models substantially lower the cost of producing many flawed design alternatives, but not the cost of sifting through these alternatives to identify and reincorporate design knowledge from the rare partial instances of success. 
In such domains, element-centric interaction brings increased \emph{decision granularity}: 
breaking down the assessment of generated designs (e.g., images) into a process of making many small decisions about individual design elements (i.e., trinkets)---rather than a process of deciding whether to reuse or discard each generated design wholesale---increases both the frequency and the specificity of designer feedback that is reincorporated into future design iterations, yielding a process that \emph{accumulates design knowledge} more efficiently as it moves forward through multiple rounds of iteration.

The value of decision granularity may also drive the increasing adoption of collage-inspired user interfaces in other genAI-supported creative domains, for instance writing~\cite{BuschekCollage}. Fragmenting generated artifacts into individually meaningful elements scaffolds decision-making at the element rather than the whole-artifact level, making it easier to accumulate and reincorporate design knowledge as the design process proceeds.




\paragraph{Limitations}
Although Study~2 introduced a four-month delay, participants revisited histories produced during bounded 30-minute Study~1 sessions. Our findings therefore characterize delayed sensemaking of relatively compact exploration histories, and longer, multi-session projects may introduce additional challenges as materials and decisions accumulate over time. Future work could examine these dynamics through longer sessions and field deployments that capture more extended creative trajectories. Study~2 also involved six of the twelve Study~1 participants who chose to return, which may shape how broadly the delayed-sensemaking findings generalize across the original participant group.

Trinketry centers on an open canvas, which gave participants flexibility in arranging evolving materials but became harder to manage as outputs accumulated---revealing a broader tension in canvas-based provenance interfaces between flexibility and scalability. Future systems might explore semi-constrained canvases with lightweight automatic layout, expandable or groupable local workspaces, or system-assisted clustering and archival to preserve traceability while reducing organizational overhead.

For semantic linking, we approximate relatedness using a multi-modal similarity function over trinket representations rather than human relevance judgments. As a result, we primarily capture relationships reflected in available visual and textual signals, and may underrepresent ones that are meaningful to designers but not easily captured by similarity alone. Future work could develop linking formulations and evaluation methods that better reflect how designers perceive relevance across evolving exploration contexts.

\section{CONCLUSION}
\label{sec:conclusion}
This work introduced trinkets, user-preserved design elements that serve as reusable materials for generation, trace anchors for navigation, and retrieval queries for prior material. Across our studies, we found that whole-image branching and element-centered traces supported different modes of visual exploration and reflection. During exploration, whole-image branching enabled faster, broader divergence, whereas Trinketry supported more deliberate, goal-directed exploration around selected elements. Four months later, whole-image histories provided a clearer global chronological overview, while trinket-centered traces supported more precise reconstruction of relationships among elements and helped cue major stages and turning points in the design process. For creators, these cues could sometimes reactivate the rationale behind earlier decisions, but such recovery was incomplete. For external reviewers, action and material traces made what happened considerably more legible than why it happened. Building on these findings, future systems could pair provenance traces with lightweight annotations of decision context, helping design histories preserve the evolving materials, actions, intentions, and reasoning that shape creative exploration.


\begin{acks}
This work was supported by the National Science and Technology Council
\end{acks}

\bibliographystyle{ACM-Reference-Format}

\bibliography{bibliography}


\appendix
\clearpage
\onecolumn
\section{Appendix A: Demographic Details}

\begin{table}[ht]
\centering
\small
\setlength{\tabcolsep}{3pt}
\renewcommand{\arraystretch}{1.1}

\begin{tabular}{
|c
|c
|c
|p{0.10\linewidth}
|p{0.36\linewidth}
|c
|c
|c|}
\hline
ID & Age & YoE & Industry & GenAI Tools Used & Formative & Study 1 & Study 2 \\
\hline
1  & 28 & 5  & Game 
& ChatGPT, Gemini 
& $\checkmark$ & $\checkmark$ & $\checkmark$ \\
\hline

2  & 28 & 4  & Film/TV, Animation 
& Midjourney, Stable Diffusion, Flux 
& $\checkmark$ & $\checkmark$ & $\checkmark$ \\
\hline

3  & 40 & 15 & Game 
& ChatGPT, Gemini 
& $\checkmark$ & $\checkmark$ & $\checkmark$ \\
\hline

4  & 32 & 9  & Film/TV 
& Midjourney, ChatGPT, Meshy, Kling, Moonland 
& $\checkmark$ & $\checkmark$ & \\
\hline

5  & 28 & 4  & Game 
& Flux 
& & $\checkmark$ & \\
\hline

6  & 27 & 3  & Animation Student 
& ChatGPT, Gemini 
& & $\checkmark$ & $\checkmark$ \\
\hline

7  & 32 & 8  & Animation 
& ChatGPT, Gemini, Moonland, Kling, Meshy, Grok, Beeble 
& & $\checkmark$ & \\
\hline

8  & 28 & 6  & Animation 
& ChatGPT 
& & $\checkmark$ & $\checkmark$ \\
\hline

9  & 40 & 10 & Game 
& Nano Banana 
& & $\checkmark$ & \\
\hline

10 & 45 & 21 & Game 
& Stable Diffusion, ChatGPT, Flux, ComfyUI, Gemini, Grok, Sora 2, Flow, Claude 
& & $\checkmark$ & \\
\hline

11 & 32 & 8  & Film/TV, Animation 
& Midjourney, ChatGPT, Nano Banana, Seedance, Gemini, Moonland 
& & $\checkmark$ & \\
\hline

12 & 31 & 7  & Game 
& ChatGPT, Gemini 
& & $\checkmark$ & $\checkmark$ \\
\hline
\end{tabular}

\caption{Demographic details and study participation of participants.}
\Description[Demographics and study participation for 12 participants.]{
Demographics table listing 12 participants with columns for participant ID,
age, years of experience, industry, generative AI tools used, and participation
in the formative study, Study 1, and Study 2. All 12 participants took part in
Study 1. Participants 1 through 4 also took part in the formative study. Six
participants---1, 2, 3, 6, 8, and 12---returned for Study 2.

Ages range from 27 to 45 years. Participants report 3 to 21 years of experience
and represent the game, film and television, and animation industries, including
one animation student. Participants had prior experience with a range of
generative AI tools, including ChatGPT, Gemini, Midjourney, Stable Diffusion,
Flux, Meshy, Kling, Moonland, Grok, Beeble, Nano Banana, ComfyUI, Sora 2, Flow,
Claude, and Seedance.
}
\label{tab:demographics}
\end{table}

\section{Multimodal Retrieval Scoring}
\label{app:retrieval-scoring}
In Trinketry, each indexed item \(t\) is represented by an image embedding set \(V_t\) and a text embedding set \(T_t\), computed using SigLIP2~\cite{tschannen2025siglip}; either representation may be absent depending on the trinket type.

To compare any two available embedding sets \(A\) and \(B\), we reduce their pairwise cosine similarities using a symmetric mean-of-maxima pooling operator \(P_{\text{smm}}\) inspired by prior late-interaction scoring methods~\cite{colbert,faysse2025colpali,VideoColBERT}:
\[
P_{\text{smm}}(A,B)
=
\frac{1}{2}
\left(
\frac{1}{|A|}\sum_{a\in A}\max_{b\in B}\cos(a,b)
+
\frac{1}{|B|}\sum_{b\in B}\max_{a\in A}\cos(a,b)
\right).
\]
This symmetric form reflects mutual correspondence between two peer items while tempering the effect of unequal embedding-set sizes.

Given a query item \(t_q\) and candidate item \(t_c\), we compute scores over the available channels \(m \in \{\mathrm{ii},\mathrm{tt},\mathrm{it},\mathrm{ti}\}\), where i and t denote image and text:
\[
\begin{aligned}
s_{\mathrm{ii}}(t_q,t_c) &= P_{\text{smm}}(V_{t_q},V_{t_c}), &
s_{\mathrm{tt}}(t_q,t_c) &= P_{\text{smm}}(T_{t_q},T_{t_c}),\\
s_{\mathrm{it}}(t_q,t_c) &= P_{\text{smm}}(V_{t_q},T_{t_c}), &
s_{\mathrm{ti}}(t_q,t_c) &= P_{\text{smm}}(T_{t_q},V_{t_c}).
\end{aligned}
\]
We then form a single cross-modal score
\[
s_{\mathrm{cross}}(t_q,t_c)=
\begin{cases}
\dfrac{1}{2}\bigl(s_{\mathrm{it}}(t_q,t_c)+s_{\mathrm{ti}}(t_q,t_c)\bigr), & \text{if both are available},\\[6pt]
s_{\mathrm{it}}(t_q,t_c), & \text{if only } s_{\mathrm{it}} \text{ is available},\\[4pt]
s_{\mathrm{ti}}(t_q,t_c), & \text{if only } s_{\mathrm{ti}} \text{ is available}.\\
\end{cases}
\]
The final retrieval channels are therefore \(\mathcal{M}(t_q,t_c)\subseteq \{\mathrm{ii},\mathrm{tt},\mathrm{cross}\}.\)

Because the modality gap causes different channels to exhibit different score scales~\cite{liang2022mind}, we rescale each available score using mismatched and matched reference means estimated from a 500-pair synthetic calibration set built from captions sampled from the MS COCO 2017 validation set~\cite{lin2014coco} and corresponding FLUX.2 [pro] generations. We then rank candidates by
\[
S(t_q,t_c)=\max_{m\in\mathcal{M}(t_q,t_c)}
\left(
\operatorname{clip}_{[0,1]}
\left(
\frac{s_m(t_q,t_c)-\mu_m^{-}}{\mu_m^{+}-\mu_m^{-}}
\right)
\right),
\]
where \(\mu_m^{-}\) and \(\mu_m^{+}\) are the mismatched and matched cosine-similarity means for channel \(m\).

In a controlled evaluation on a deterministic 60/40 held-out split of the same 500-pair COCO--FLUX.2 [pro] dataset (60\% evaluation, 40\% calibration), we paired each full-caption text trinket with an image trinket built from the same generated image and caption keyword chunks. Compared with no rescaling, the full method improved Recall@1 from 0.595 to 0.912, Recall@5 from 0.774 to 0.995, and MRR from 0.676 to 0.945, supporting our choice to rescale channel scores before computing the final retrieval score.

\section{Trace Co-Review Interview Protocol}
\label{Supplement_Trace_CoReview}
\small\ttfamily
\begin{lstlisting}
Part 1: Process Review
- Briefly walk me through how this design evolved.
- Where do you think the process really began to take shape?
- Where do you think your idea started to converge?

Part 2: Turning Points & Idea Sources
- Where did you make an important change in direction, and why?
- Which earlier ideas became important later?
- [Optional] Do you see earlier ideas that resurfaced later in a different form?
- [Optional] Were there branches you abandoned and later returned to?

Part 3: Element-Level Reasoning
- Where did this important element first appear?
- Was there any idea you wanted to use but could not successfully employ? Why?
- [Optional] Can you give examples of which parts were preserved, changed, 
or recombined over time?

Part 4: Reflection & Metacognition
- What would you revisit if you continued this session?
- What does this trace reveal about your process that you did not notice during the task?
- [Optional] Can you identify a moment where the system drifted from your intent? 
How did you deal with that? Why?
\end{lstlisting}

\subsection{Adaptations for Study 2 Delayed Self-Sensemaking}
Study 2 retained the core questions concerning process structure,
turning points, element provenance, reuse, and abandoned material.
Because participants revisited their work after an approximately
four-month delay, we shifted the additional probes from immediate
tracing experience toward delayed memory and recoverability.

\small\ttfamily
\begin{lstlisting}
Additional delayed-sensemaking probes:
- What parts of the process remain particularly memorable?
- What parts are now uncertain or difficult to recover?
- How confident are you that you capture what mattered most in your process?
\end{lstlisting}

\end{document}